\documentclass[12pt]{article}

\pdfoutput=1 
\usepackage{jheppub}

\usepackage{graphicx} 
\usepackage{epsfig} 
\usepackage{epstopdf} 
\usepackage{amsmath} 
\usepackage{amsfonts} 
\usepackage{amssymb} 
\usepackage{ccaption}

\captionnamefont{\bfseries} \captiontitlefont{\small\sffamily} \captiondelim{: } \hangcaption

\def\bulkD{{\tilde D}} 
\def\bulkJ{{\tilde J}} 
\def\bdyI{I}
\def\bulkI{{\tilde I}}

\def\splrel{\asymp} 
\def\splsep{{\cal S}} 
\def\bulksplsep{{\tilde {\cal S}}}

\def\AdS#1{\AdS$_{#1}$}

\def\bulk{{\cal M}} 
\def\bdy{{\cal B}} 
\def\overM{\tilde \bulk}
\def\bulkC{{\bar \bulk}}
\def\regA{{\cal A}} 
\def\regAc{{\cal A}^c} 
\def\rhoA{{\rho_{\regA}}} 
 
\def\entsurf{
\partial \regA} 
\def\domdA{D[\regA]} 
\def\domdAc{D[\regAc]} 
\def\domd#1{D[#1]}
\def\extr{{\cal E}_\regA} 
\def\CIS{{\Xi_\regA}} 
\def\homsurfA{{\cal R}_{\regA}} 
\def\homsurfAc{{\cal R}_{\regA}^c} 
\def\CWA{{\cal W}_{\cal C}[\regA]} 
\def\CWAc{{\cal W}_{\cal C}[{\regAc}]} 
\def\EWA{{\cal W}_{\cal E}[\regA]} 
\def\EWAc{{\cal W}_{\cal E}[\regAc]}
\def\shadow{{\cal Q}}
 
\def\AdS#1{AdS$_{#1}$}
\def\SAdS#1{Schwarzschild-AdS$_{#1}$}

\def\ket#1{\mid \! #1\rangle}

\def\be{\begin{equation}}
\def\ee{\end{equation}}
\def\RR{{\mathbb R}}

\DeclareMathOperator{\Tr}{Tr}

\newtheorem{theorem}{Theorem}
\newtheorem{lemma}[theorem]{Lemma}

\newtheorem{corollary}[theorem]{Corollary}

\definecolor{rust}{rgb}{0.8,0.2,0.2} 
\definecolor{purple}{rgb}{0.8,0.1,0.9} 
\definecolor{olivegreen}{rgb}{0,0.52,0.17}

\title{Causality \& holographic entanglement entropy}

\author{Matthew Headrick$^a$,}
\author{Veronika E. Hubeny$^b$,}
\author{Albion Lawrence$^a$,}
\author{ \\  Mukund Rangamani$^b$}

\affiliation[a]{ Martin Fisher School of Physics, Brandeis University, \\
MS 057, 415 South Street, Waltham, MA 02454, USA.}
\affiliation[b]{
Centre for Particle Theory \& Department of Mathematical Sciences,\\
Science Laboratories, South Road, Durham DH1 3LE, UK.}
\emailAdd{headrick@brandeis.edu}
\emailAdd{veronika.hubeny@durham.ac.uk}
\emailAdd{albion@brandeis.edu}
\emailAdd{mukund.rangamani@durham.ac.uk}

\abstract{
We identify conditions for the entanglement entropy as a function of spatial region to be compatible with causality in an arbitrary relativistic quantum field theory. 
We then prove that the covariant holographic entanglement entropy prescription (which relates entanglement entropy of a given spatial region on the boundary to the area of a certain extremal surface in the bulk) obeys these conditions, as long as the bulk obeys the null energy condition.
While necessary for the validity of the prescription, this consistency requirement is quite nontrivial from the bulk standpoint, and therefore provides important additional evidence for the prescription.
In the process, we introduce a codimension-zero bulk region, named the entanglement wedge, naturally associated with the given boundary spatial region.
We propose that the entanglement wedge is the most natural bulk region corresponding to the boundary reduced density matrix.
} 

\preprint{DCPT-14/33, BRX-TH-6284}


\begin{document}
\maketitle

\section{Introduction} \label{sec:intro}


One of the remarkable features of the holographic AdS/CFT correspondence is the geometrization of quantum-field-theoretic concepts. While certain aspects of recasting field-theory quantities into geometric notions have been ingrained in our thought, we are yet to fully come to grips with new associations between QFT and bulk geometry. A case in point is the fascinating connection of quantum entanglement and  spacetime geometry. 
The genesis of this intricate and potentially deep connection harks back to the observation of Ryu-Takayanagi (RT) \cite{Ryu:2006ef,Ryu:2006bv} and subsequent covariant generalization by Hubeny-Rangamani-Takayanagi (HRT) 
\cite{Hubeny:2007xt}  that the entanglement entropy of a quantum field theory is holographically computed by the area of a particular extremal surface in the bulk. In recent years, much effort has been expended in trying to flesh out the physical implications of these constructions and in promoting the geometry/entanglement connection to a deeper level \cite{Swingle:2009bg, VanRaamsdonk:2009ar,VanRaamsdonk:2010pw, Maldacena:2013xja} which can be summarized rather succinctly in terms of the simple phrases ``entanglement builds bridges'' and ``ER = EPR''.
Whilst any connection between entanglement and geometry is indeed remarkable, further progress is contingent on the accuracy and robustness of this entry in the holographic dictionary. Let us therefore take stock of the status quo.\footnote{ We will focus exclusively on local QFTs with conformal UV fixed points  which are holographically dual to asymptotically AdS spacetimes in 
{\em two-derivative} theories of gravity.}

The RT proposal is valid for static states of a holographic field theory, which allows one to restrict attention to a single time slice ${\tilde \Sigma}$ in the bulk spacetime $\bulk$. 
The entanglement entropy of a region $\regA$ on the corresponding
Cauchy slice $\Sigma$ of the boundary spacetime $\bdy$ is computed by the area of a certain bulk minimal surface which lies on ${\tilde \Sigma}$.
In this case we have a lot of confidence in this entry to the AdS/CFT dictionary; firstly the RT formula obeys rather non-trivial general properties of entanglement entropies such as strong subadditivity \cite{Headrick:2007km,Hayden:2011ag,Headrick:2013zda}, and secondly a general argument has been given for it in the context of Euclidean quantum gravity \cite{Lewkowycz:2013nqa}.

However, it should be clear from the outset that restricting oneself to static states is overly limiting.
Not only is the field theory notion of entanglement entropy valid in a broader, time-dependent, context, but more importantly,
one cannot hope to infer all possible constraints on the holographic map without considering time dependence.

The HRT proposal, which generalizes the RT construction to arbitrary time-de\-pend\-ent configurations by promoting a minimal surface on  ${\tilde \Sigma}$ to an {\it extremal} surface $\extr$ in $\bulk$, allows one to confront geometric questions in complete generality. However, this proposal has passed far fewer checks, and an argument deriving it from first principles is still lacking. This presents a compelling opportunity to test the construction against field-theory expectations and see how it holds up. Since the new ingredient in HRT  is time-dependence, the crucial property to check is causality. The present discussion therefore focuses on verifying that {\it the HRT prescription is consistent with field-theory causality}.\footnote{ As we elaborate in the course of our discussion this result follows from Theorem 6 of \cite{ Wall:2012uf}. As this is however not widely appreciated we focus on proving the result from a different perspective highlighting certain novel bulk constructs in the process.}

Let us start by considering the implications of CFT causality on entanglement entropy, in order to extract the corresponding requirements to be upheld by its putative bulk dual. As we will explain in detail in \S\ref{sec:overview}, there are two such requirements. First, the entanglement entropy is a so-called \emph{wedge observable}. This means that two spatial regions $\regA$, $\regA'$ that share the same domain of dependence, $D[\regA]=D[\regA']$, have the same entanglement entropy, $S_\regA=S_{\regA'}$; this follows from the fact that the corresponding reduced density matrices $\rho_\regA$, $\rho_{\regA'}$ are unitarily related \cite{Casini:2003ix}. Second, fixing the initial state, a perturbation to the Hamiltonian with support contained entirely inside $D[\regA]\cup D[\regA^c]$ (where $\regA^c$ is the complement of $\regA$ on a Cauchy slice) cannot affect $S_\regA$. The reason is that we can choose a Cauchy slice $\Sigma'$ that lies to the past of the support and contains a region $\regA'$ with $D[\regA']=D[\regA]$; since the perturbation cannot change the state on $\Sigma'$, it cannot affect $S_{\regA'}$, which by the previous requirement equals $S_\regA$. Time-reversing the argument shows that, similarly, $S_\regA$ cannot be affected by a perturbation in $D[\regA]\cup D[\regA^c]$ when we consider time evolution toward the past with a fixed final state.

Having specified the implications of causality for the entanglement entropy in the field theory, let us now translate them into requirements on its holographic dual. First, in order to ensure that the HRT formula in general gives the same entanglement entropy for $\regA$ and $\regA'$, they should have the same extremal surface, $\extr={\cal E}_{\regA'}$. Second, in order for $\extr$ to be safe from influence by perturbations of the boundary Hamiltonian in $\domdA$ and $D[\regA^c]$ (when evolving either toward the future or toward the past), it has to be causally disconnected from those two regions. This means that the extremal surface has to lie in a region which we dub the {\it causal shadow}, denoted by $\shadow_{\entsurf}$ and defined in \eqref{shadowdef} as the set of bulk points which are spacelike-separated from $\domdA\cup\domdAc$.

This causality requirement takes an interesting guise in the case where $\regA$ is an entire Cauchy slice for a boundary.  If this is the only boundary, and the bulk is causally trivial, then there is no causal shadow; indeed, $\extr=\emptyset$, corresponding to the fact that the entanglement entropy of the full system vanishes in a pure state. However, if the state is {\it not} pure, the bulk geometry is causally nontrivial: typically the bulk black-hole spacetime has two boundaries, dual to two field theories in an entangled state (which can be thought of as purifying the thermal state of the theory on one boundary). If we take the region $\regA$ to be a Cauchy slice for one boundary and $\regAc$ a Cauchy slice for the other, then the extremal surface whose area, according to HRT, measures the amount of entanglement between the two field theories must lie in a region out of causal contact with either boundary.\footnote{ For the well-known eternal static Schwarzschild-AdS case, the shadow region degenerates to the bifurcation surface, but we will see that in general it is a finite codimension-zero bulk region.}

How trivial or expected is the claim that the extremal surface resides in the causal shadow? 
It is interesting to note that for {\it local} CFT observables, analogous causality violation is in fact disallowed by the gravitational time-delay theorem of Gao and Wald \cite{Gao:2000ga}. This theorem, which assumes that the bulk satisfies the null energy condition, implies that a signal from one boundary point to another cannot propagate faster through the bulk than along the boundary, ensuring that bulk causality respects boundary causality. However, since entanglement entropy is a more nonlocal quantity, which according to HRT is captured by a bulk surface that can go behind event and apparent horizons \cite{Hubeny:2002dg,AbajoArrastia:2010yt} and penetrate into causally disconnected regions from the boundary, it is far less obvious whether CFT causality will survive in this context.  

Let us first consider a static example.  Although it is guaranteed to be consistent with CFT causality since it is covered by the RT prescription which is ``derived'' from first principles, it is useful to gain appreciation for how innocuous or far-fetched causality violation would appear in the more general case.
Intriguingly, already the simplest case of pure AdS reveals the potential for things to go wrong. 
\begin{figure}
\begin{center}
\includegraphics[width=2in]{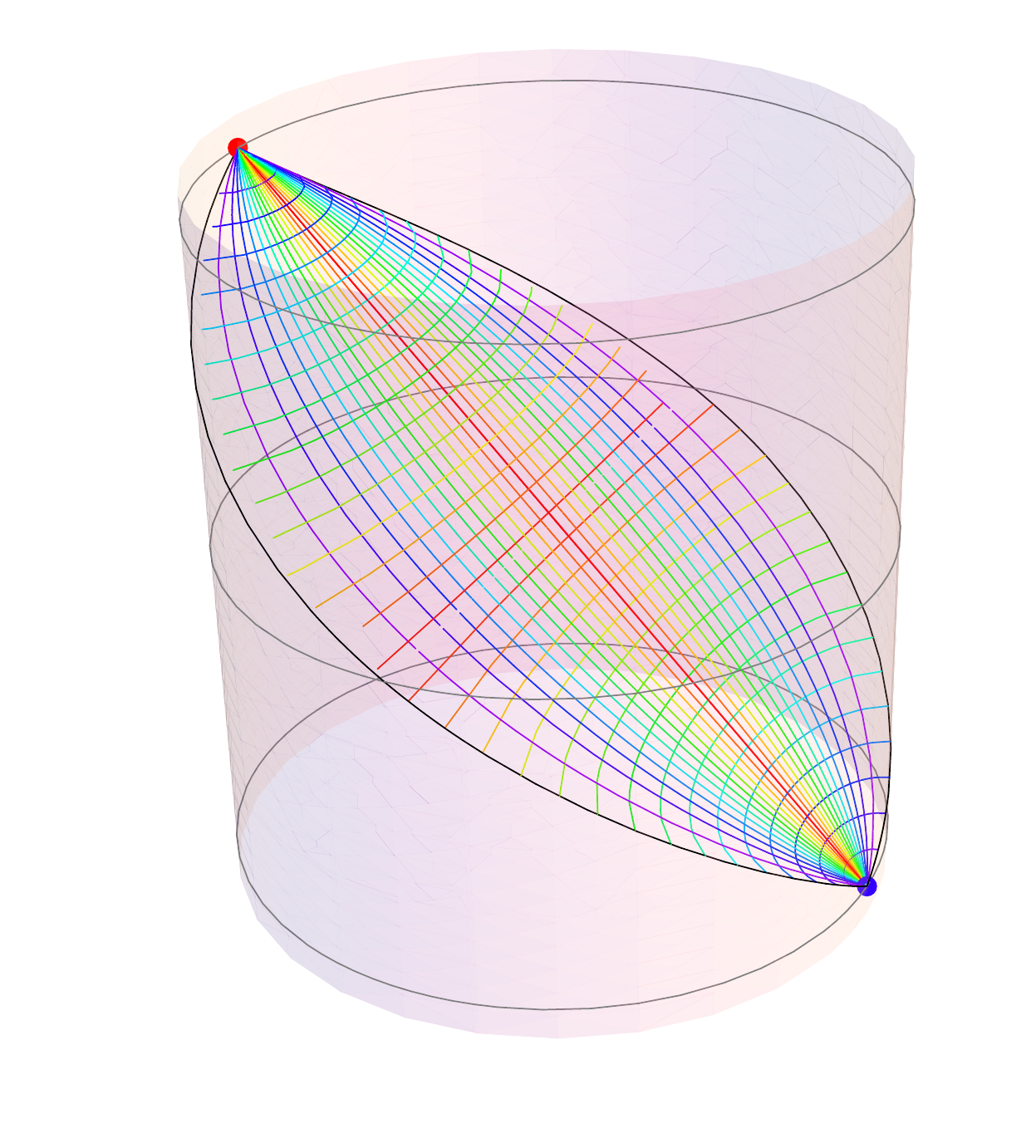}
\caption{
For \AdS{3}, the RT formula satisfies field-theory causality marginally. The plane generated by null geodesics (color-coded by angular momentum) from a given boundary point (blue) is also ruled by spacelike geodesics at constant time (color-coded by time).
}
\label{f:AdS3geods}
\end{center}
\end{figure}
As illustrated in Fig.\ \ref{f:AdS3geods}, the null congruence from a single boundary point (which bounds the bulk region which a boundary source at that point can influence) is simultaneously foliated by spacelike geodesics $\{ \extr \}$. So a signal that can influence a given extremal surface $\extr$ in that set can also influence $\entsurf$, thereby upholding CFT causality. However, note that here causality was maintained marginally: if the extremal surface was deformed away from $\regA$ by arbitrarily small amount, one would immediately be in danger of causality violation.

Another, less trivial, test case is the static eternal Schwarzschild-AdS black hole.  The extremal surface that encodes  entanglement between the two boundaries is the horizon bifurcation surface. Again, arbitrarily small deformation of this surface would shift it into causal contact with at least one of the boundaries, thereby endangering causality; in particular, entanglement entropy for one CFT should not be influenced by deformations in the other CFT.
For static geometries we're in fact safe because extremal surfaces do not penetrate event horizons \cite{Hubeny:2012ry}; however this is no longer the case in dynamical situations \cite{Hubeny:2002dg,AbajoArrastia:2010yt,Hartman:2013qma,Liu:2013iza,Hubeny:2013dea}. 
Moreover, as illustrated in \cite{Hubeny:2013hz}, in Vaidya-AdS geometry,  $\extr$ can be null-related to the past tip of $\domdA$,  thereby again upholding causality just marginally---an arbitrarily small outward deformation of the extremal surface would render it causally accessible from $\domdA$.
These considerations demonstrate that the question of whether the HRT prescription is consistent with field-theory causality is a highly nontrivial one.  

The main result of this paper is a proof that, if the bulk spacetime metric obeys the null energy condition, then the extremal surface $\extr$ does indeed obey both of the above requirements. We conclude that the HRT formula is consistent with field-theory causality. This theorem can be viewed as a generalization of the Gao-Wald theorem \cite{Gao:2000ga}. We regard it as a highly nontrivial piece of evidence in favor of the HRT formula. Along the way, we will also slightly sharpen the statement of the HRT formula, and in particular clarify the homology condition on $\extr$. 

Partial progress towards this result was achieved in \cite{Hubeny:2012wa, Hubeny:2013gba}, which showed that the extremal surface $\extr$ generically lies outside of the ``causal wedge" of $\domdA$, the intersection of the bulk causal future and causal past of $\domdA$. (However, these works did not make the connection to field-theory causality). A stronger statement equivalent to our theorem was proved in \cite{Wall:2012uf} (cf., Theorem 6) and it is noted in passing that this would ensure field theory causality. We present an alternate proof which brings out some of of the bulk regions more cleanly and make the connections with boundary causality more manifest.

As a byproduct of our analysis, we will identify a certain bulk spacetime region, which we call the \emph{entanglement wedge} and denote $\EWA$, which is bounded on one side by $D[\regA]$ and on the other by $\extr$. Apart from providing a useful quantity in formulating and deriving our results, the entanglement wedge is, as we will argue, the bulk region most naturally associated with the boundary reduced density matrix $\rhoA$.

The outline of this paper is as follows. We begin in \S\ref{sec:overview} with an overview of the causal domains of interest on each side of the gauge-gravity duality, and motivate and state the core theorem of the paper, which shows that the HRT proposal is consistent with boundary causality.  We motivate one of the major implications of our theorem by considering spherically symmetric deformations of the eternal black hole containing a region out of causal contact with both asymptotically AdS boundaries, the \emph{causal shadow}, and showing that the HRT surface lies in this causal shadow. In \S\ref{sec:examples}, we begin to develop some intuition used in the proof of our main theorem, by considering classes of null geodesic congruences in \AdS{3}.  In \S\ref{sec:results} we prove the general theorem which establishes the main result of the paper. We conclude in \S\ref{sec:discuss} with a discussion of the physical implications of our result and open questions.

\bigskip
\noindent
{\em Note added:} While this paper was nearing completion \cite{Engelhardt:2014gca} appeared on the arXiv, which has some overlap with the present work.
It introduces the notion of quantum extremal surfaces and argues that for bulk theories that satisfy the generalized second law such surfaces satisfy the causality constraint.

\section{Causal domains and entanglement entropy}
\label{sec:overview}
In this section we will state our basic results and discuss some of their implications.  The specific proof, and some additional results, will be presented in \S\ref{sec:results}.  In \S\ref{sec:discuss} we will suggest some further interpretations of our results, particularly regarding the dual of the reduced density matrix.

We will open in \S\ref{sec:cqft} by deriving the causality properties of entanglement entropy in a QFT, and setting up some notation regarding causal domains which will be useful in the sequel.  In \S\ref{sec:cbulkexp}, we will review the HRT formula and discuss various causal regions in the bulk.  In \S\ref{sec:bulkextc}, we state the basic theorem and some implications for the bulk causal structure relative to specific regions arising in the HRT conjecture. \S\ref{sec:gexpt} spells out a particular consequence of our results for spacetimes with multiple boundaries.

Where left unspecified, our notation follows \cite{Wald:1984ai}.

\subsection{Causality of entanglement entropy in QFT}
\label{sec:cqft}

Consider a local quantum field theory (QFT) on a $d$-dimensional globally hyperbolic spacetime $\bdy$.  The state on a given Cauchy slice\footnote{ \label{Cauchydef}Throughout this paper we will require all Cauchy slices to be \emph{acausal} (no two points are connected by a causal curve). This is slightly different from the standard definition in the general-relativity literature, in which a Cauchy slice is merely required to be achronal. The reason is to ensure that different points represent independent degrees of freedom, which is useful when we decompose the Hilbert space according to subsets of the Cauchy slice.} $\Sigma$ is described by a density matrix $\rho_\Sigma$; this could be a pure or mixed state.
We are interested in the entanglement 
between the degrees of freedom in a region\footnote{ Technically, $\regA$ is defined as the interior of a codimension-zero submanifold-with-boundary in $\Sigma$, $\partial\regA$ is the boundary of that submanifold, and $\regAc:=\Sigma\setminus(\regA\cup\partial\regA)$.} 
 $\regA \subset \Sigma$ and its complement $\regAc$.  Following established terminology, we call the boundary  $\entsurf$ the {\it entangling surface}. 

The entanglement entropy is defined by first decomposing the Hilbert space ${\cal H}$ of the QFT into ${\cal H}_{\regA} \otimes {\cal H}_{\regAc}$, after imposing some suitable cutoff.\footnote{ In the case of gauge fields, this decomposition is not possible even on the lattice.  Instead, one must extend the Hilbert spaces ${\cal H}_{\regA}$, ${\cal H}_{\regAc}$ to each include degrees of freedom on $\entsurf$, so that ${\cal H} \subset {\cal H}_{\regA} \otimes {\cal H}_{\regAc}$ \cite{Buividovich:2008gq,Donnelly:2011hn,Casini:2013rba,Donnelly:2014gva}.} The reduced density matrix $\rho_{\regA} := \Tr_{{\cal H}_{\regAc}} \rho_\Sigma$ captures the entanglement between $\regA$ and $\regAc$; in particular, the entanglement entropy is given by its von Neumann entropy: $S_{\regA} := - \Tr\left( \rho_{\regA} \ln \rho_{\regA}\right)$. For holographic theories, we expect that this quantity has good properties in the large-$N$ limit,\footnote{ Technically, by ``large-$N$" we mean large $c_\text{eff}$, where 
$c_\text{eff} $ is a general count of the degrees of freedom (see \cite{Marolf:2013ioa} for the general definition of $c_\text{eff}$).} unlike the R\'enyi entropies $S_{n,\regA} := - \frac{1}{n-1} \ln \Tr \left(\rho_{\regA}^n\right)$ \cite{Headrick:2010zt,Headrick:2013zda}. Note that both quantities are determined by the eigenvalues of $\rhoA$, and are thus insensitive to unitary transformations of $\rhoA$.

Now, since $\Sigma$ is a Cauchy slice, the future (past) evolution of initial data on it allows us to reconstruct the state of the QFT on the entirety of ${\cal B}$. In other words, the past and future domains of dependence of $\Sigma$ , $D^\pm[\Sigma]$, together make up the background spacetime on which the QFT lives, i.e., 
$D^+[\Sigma] \cup D^-[\Sigma] = \bdy$. 
Likewise,  the domain of dependence of $\regA$,  $\domdA = D^+[\regA] \cup D^-[\regA]$,  is the region where the reduced density matrix $\rhoA$ can be uniquely evolved once we know the Hamiltonian acting on the reduced system in $\regA$.\footnote{ We remind the reader that $D[\regA]$ is defined as the set of points in $\bdy$ through which every inextendible causal curve intersects $\regA$. Note that, given that we have defined $\regA$ as an open subset of $\Sigma$, $D[\regA]$ is open subset of $\bdy$.}

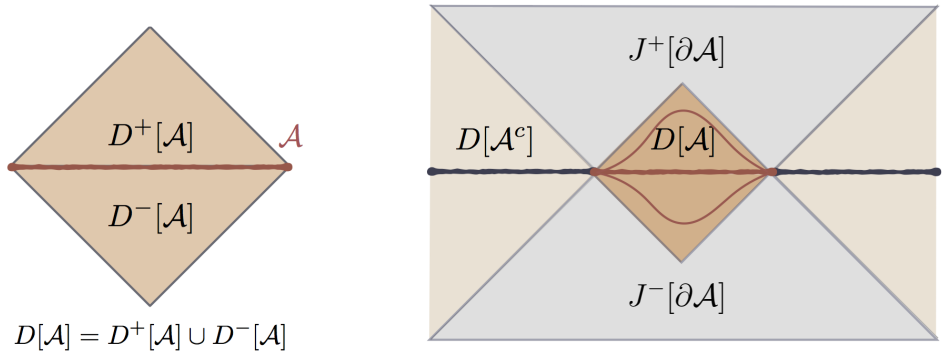
\begin{figure}
\begin{center}
\includegraphics[width=5in]{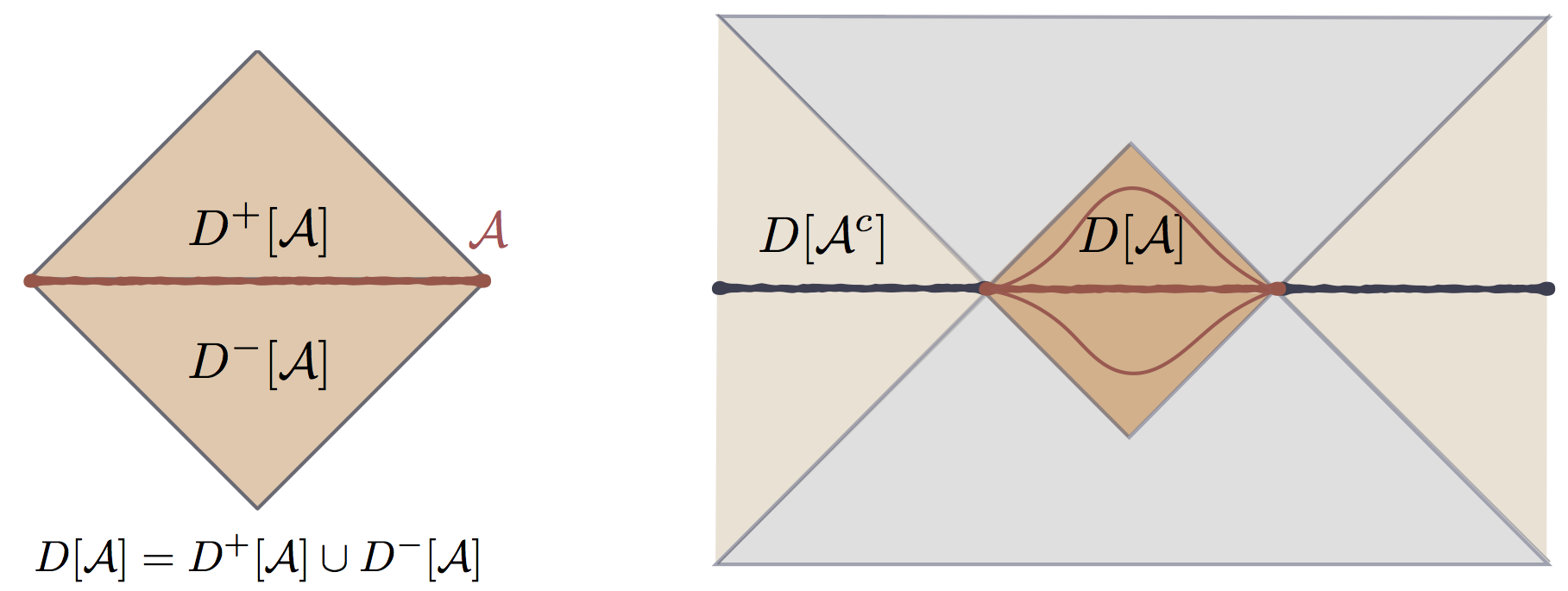}
\setlength{\unitlength}{0.1\columnwidth}
\begin{picture}(0.3,0.4)(0,0)
\put(-2.5,2.6){\makebox(0,0){$J^+[\entsurf]$}}
\put(-2.5,0.5){\makebox(0,0){$J^-[\entsurf]$}}
\end{picture}
\caption{
An illustration of the causal domains associated with a region $\regA$, making manifest the decomposition of the spacetime into the four distinct domains indicated in \eqref{bdy4d}.
Two deformations $\regA'$ are also included for illustration in the right panel.
}
\label{f:bdy4d}
\end{center}
\end{figure}

$\regAc$ similarly has its domain of dependence $\domdAc$. However, unless $\regA$ comprises the entire Cauchy slice, the two domains do not make up the full spacetime, $\domdA \cup \domdAc\neq \bdy$, since we have to account for the regions which can be influenced by the entangling surface $\entsurf$. Denoting the causal  future (past) of a point $p\in {\cal B}$ by $J^\pm(p)$ 
we find that we have to keep track of the regions $J^\pm[\entsurf]$ which are not contained in either $\domdA$ or $\domdAc$. As a result, the  full spacetime $\bdy$ decomposes into four causally-defined regions: the domains of dependence of the region and its complement, and the causal future and past of the entangling surface:
\begin{equation}
\bdy = \domdA \cup \domdAc \cup J^+[\entsurf] \cup J^-[\entsurf]\,.
\label{bdy4d}
\end{equation}	
These four regions are non-overlapping (except that $J^\pm[\partial\regA]$ both include $\partial\regA$). See Fig.~\ref{f:bdy4d} for an illustration of this decomposition. 
Although this decomposition is fairly obvious pictorially, for completeness we provide a proof in \S\ref{sec:results} (cf.\ theorem \ref{decomposition}).

The decomposition \eqref{bdy4d} is particularly convenient for formulating the QFT causality constraint. Recall that the eigenvalues of the reduced density matrix $\rhoA$, and hence the R\'enyi and von Neumann entropies, are invariant under unitary transformations which act on ${\cal H}_{\regA}$ alone or on ${\cal H}_{\regAc}$ alone.  These include perturbations of the Hamiltonian and local unitary transformations supported in the domains $D[\regA]$ or $D[\regAc]$. 
In particular, if we consider another region $\regA'$ of a Cauchy slice $\Sigma'$ such that $D[\regA]=D[\regAc]$ (as indicated in Fig.~\ref{f:bdy4d}), then the state $\rho_{\Sigma'}$ is related by a unitary transformation to the state $\rho_\Sigma$.  It is clear that such a transformation can be constructed from operators localized in $\regA$, and so does not change the entanglement spectrum of $\rhoA$. Furthermore, if we fix the state at $t\to-\infty$, then a perturbation to the Hamiltonian with support $R$ cannot affect the state on a Cauchy slice to the past of $R$ (i.e.\ that doesn't intersect $J^+[R]$). Such a perturbation can therefore affect the entanglement spectrum only if $R$ intersects $J^-[\entsurf]$, because otherwise we can imagine evaluating $S_\regA$ by using a sufficiently early Cauchy slice $\Sigma' \supset \entsurf$ that passes to the past of $R$. Similarly, if we fix the state at $t\to+\infty$, the spectrum can be affected only by perturbations in $J^+[\entsurf]$. In summary, we have the following properties of $\rhoA$: 
\begin{itemize}
\item 
The entanglement spectrum of $\rhoA$ depends only on the domain $D[\regA]$ and not on the particular choice of Cauchy slice $\Sigma$. The spectrum is thus a so-called ``wedge observable'' (although it is not, of course, an observable in the usual sense).
\item 
Fixing the state in either the far past or the far future, the entanglement spectrum of $\rhoA$ is insensitive to any local deformations of the Hamiltonian in $\domdA$ or $\domdAc$.
\end{itemize}
These are the crucial causality requirements that entanglement (R\'enyi) entropies are required to satisfy in any relativistic QFT.

 The essential result of this paper is that the HRT proposal  for computing $S_{\regA}$ satisfies these causality constraints. In the conclusions we will revisit the question of what the dual of $\rhoA$, and thus of the data in $D[\regA]$, might be.

\subsection{Bulk geometry and holographic entanglement entropy}
\label{sec:cbulkexp}

Let us now restrict attention to the class of holographic QFTs, which are theories dual to classical dynamics in some bulk asymptotically AdS spacetime. To be precise, we only consider strongly coupled QFTs in which the classical gravitational dynamics truncates to that of Einstein gravity, possibly coupled to matter which we will assume satisfies the null energy condition. 

The dynamics of the QFT on $\bdy$ is described by classical gravitational dynamics on a bulk 
asymptotically locally AdS spacetime $\bulk$ with conformal boundary $\bdy$, the spacetime where the field theory lives. We define $\overM:=\bulk\cup\bdy$. $\overM$ is endowed with a metric $\tilde g_{ab}$ which is related by a Weyl transformation to the physical metric $g_{ab}$ on $\bulk$, $\tilde g_{ab}=\Omega^2g_{ab}$, where $\Omega\to0$ on $\bdy$.\footnote{ These are necessary but not sufficient conditions for the spacetime to be asymptotically AdS.}   Causal domains on $\overM$ will be denoted with a tilde to distinguish them from their boundary counterparts, e.g., $\bulkJ^\pm(p)$ will denote the causal future and past of a point $p$ in $\overM$ and $\bulkD[R]$ will denote the domain of dependence of some set $R\subset\overM$.

It will also be useful to introduce a compact notation to indicate when  two points $p$ and $q$ are spacelike-separated; for this we adopt the notation $\splrel$, i.e.\
\begin{equation}
p\splrel q \;\; \Leftrightarrow \;\;\text{$\nexists$ a causal curve between $p$ and $q$.}
\label{splrel}
\end{equation}	
Moreover, to denote regions that are spacelike separated from a point, we will use $\splsep(p)$ and $\bulksplsep(p)$  in the boundary and bulk respectively,
\begin{equation}
\splsep(p)  := \{ q \mid p\splrel q  \}
=\left( J^+(p) \cup J^-(p) \right)^c
\qquad {\rm and} \qquad
\bulksplsep(p)  := \left( \bulkJ^+(p) \cup \bulkJ^-(p) \right)^c\,.
\label{splregions}
\end{equation}	
Just as for other causal sets, we can extend these definitions to any region $R$, namely $\splsep[R] := \cap_{p \in R} \splsep(p)$ is the set of points which are causally disconnected from the entire region $R$, etc.

Having established our notation for general causal relations, let us now specify the notation relevant for holographic entanglement entropy. As before we will fix a region $\regA$ on the boundary. The HRT proposal \cite{Hubeny:2007xt} states that the entanglement entropy $S_\regA$ is holographically computed  by the area of a bulk codimension-two extremal surface $\extr$ that is anchored on $\entsurf$; specifically,
\begin{equation}
S_\regA = \frac{\text{Area}(\extr)}{4 G_N}\,.
\label{SAdef}
\end{equation}	
In the static (RT) case, it is known that the extremal surface is required to be homologous to $\regA$, meaning that there exists a bulk region $\homsurfA$ such that $\partial\homsurfA=\regA\cup\extr$. So far, it has not been entirely clear what the correct covariant generalization of this condition is. In particular, should it merely be a topological condition, or should one impose geometrical or causal requirements on $\homsurfA$, for example, that it be spacelike? (A critical discussion of the issues involved can be found in \cite{Hubeny:2013gta}.) In this paper, we will show that a clean picture, consistent with all aspects of field-theory causality, is obtained by requiring that $\homsurfA$ be a region of a bulk Cauchy slice.\footnote{ Technically, similarly to $\regA$, we define $\homsurfA$ to be the interior of a codimension-zero submanifold-with-boundary of a Cauchy slice $\tilde\Sigma$ of $\overM$ (with $\tilde\Sigma\cap\bdy=\Sigma$). Since $\tilde\Sigma$ itself has a boundary (namely its intersection with $\bdy$), the interior of a subset (in the sense of point-set topology) includes the part of its boundary along $\bdy$. Thus, $\homsurfA$ includes $\regA$ (but not $\extr$).}  We will call this the ``spacelike homology'' condition.\footnote{ If there are multiple extremal surfaces obeying the spacelike homology condition, then we are to pick the one with smallest area. However, in this paper we will not use this additional minimality requirement; all our theorems apply to any spacelike-homologous extremal surface.}

The homology surface $\homsurfA$ naturally leads us to the key construct pertaining to entanglement entropy,  which we call the {\em entanglement wedge} of $\regA$, denoted
by\footnote{ While we have associated it notationally with the region $\regA$, it depends only on $D[\regA]$. } $\EWA$. This can be defined as a causal set, namely the bulk domain of dependence of $\homsurfA$,
\begin{equation}
\EWA  := \bulkD[\homsurfA]\,.
\label{ewedge}
\end{equation}	
Note that the entanglement wedge is a bulk codimension-zero spacetime region, which can be equivalently identified with the region defined by the set of bulk points which are spacelike-separated from $\extr$ and connected to $\domdA$.  The latter definition has the advantage of absolving us of having to specify an arbitrary  homology surface $\homsurfA$ rather than just $\extr$ and $\domdA$.
As we shall see below, the bulk spacetime can be naturally decomposed into four regions analogously to the boundary decomposition \eqref{bdy4d}; the entanglement wedge is then the region associated with (and ending on) $\domdA$.

While we have focused on the regions in the bulk which enter the holographic entanglement entropy constructions, we pause here to note two other causal constructs that can be naturally associated with $\regA$. First of all we have the {\em causal wedge} $\CWA$ which is set of all bulk points which can both send signals to and receive signals from boundary points contained in $\domdA$, i.e.,\footnote{ Following \cite{Hubeny:2012wa}, we can also define a particular bulk codimension-two surface $\CIS$, the causal information surface, to be the rim of the causal wedge; in fact, it is the minimal area codimension-two surface lying on $\partial\CWA$. 
}
\begin{equation}
\CWA := \bulkJ^+\big[\domdA\big] \cap \bulkJ^- \big[\domdA\big].
\label{cwedge}
\end{equation}	
(The entanglement wedge $\EWA$ and causal wedge $\CWA$ are in fact special cases of the ``rim wedge'' and ``strip wedge'' introduced recently in \cite{Hubeny:2014qwa} as bulk regions associated with residual entropy.)

\begin{figure}
\begin{center}
\includegraphics[width=3in]{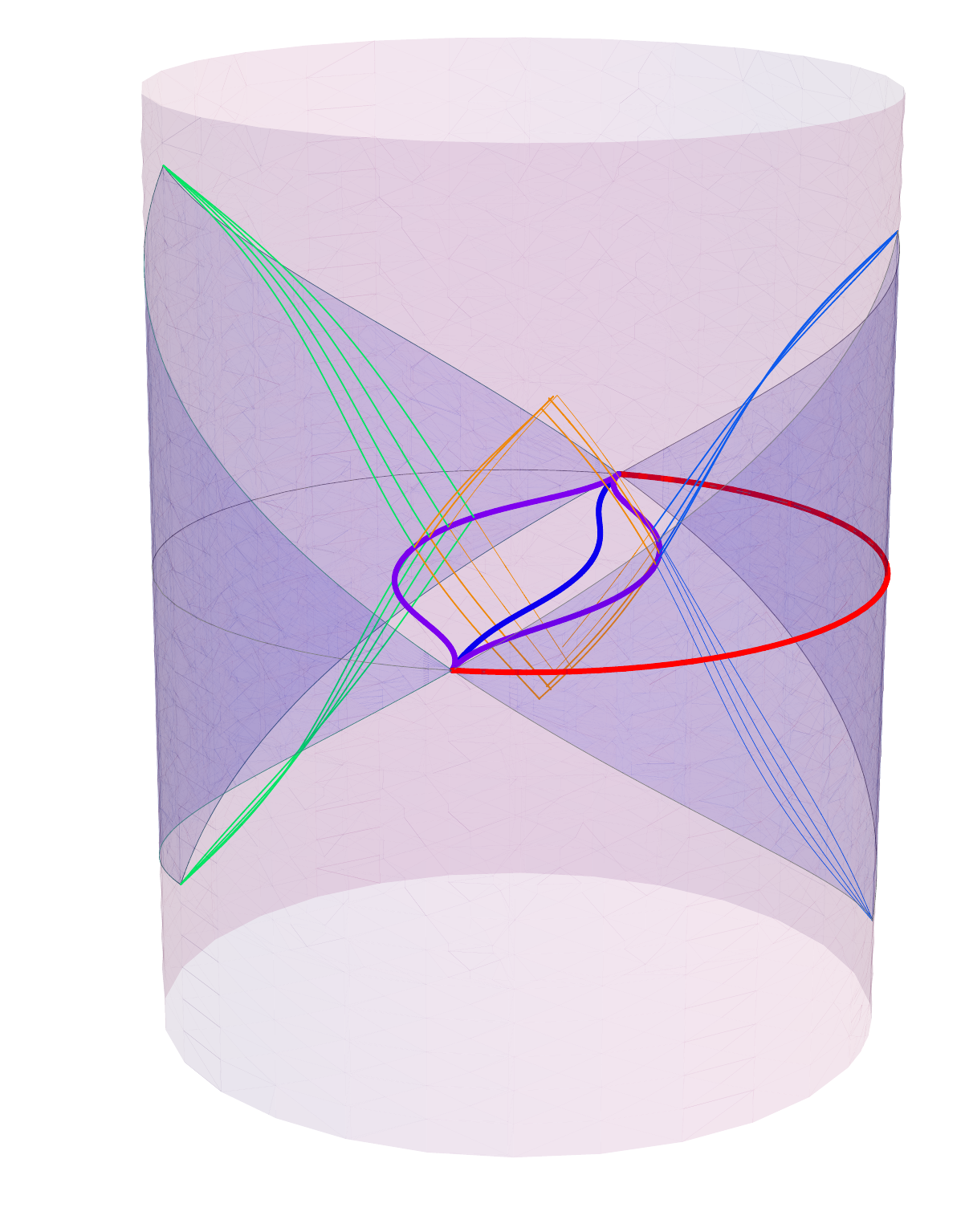}
\begin{picture}(0,0)
\setlength{\unitlength}{1cm}
\put (-3.8,5.2) {$\extr$}
\put (-0.7,5) {$\color{red}{\regA}$}
\put (-7.1,5.1) {$\regAc$}
\put (-0.7,6) {$\leftarrow\domdA$}
\put (-8.2,6.2) {$\domdAc\rightarrow$}
\put (-3.5,6.6) {$\big\downarrow$}
\put(-3.6,7.1){$ \shadow_{\entsurf}$}
\end{picture}
\caption{
Example of a causally trivial spacetime and a boundary region $\regA$ whose causal shadow is a finite spacetime region. We have engineered an asymptotically \AdS{3} geometry sourced by matter satisfying the null energy condition (see footnote \ref{fn:metric}) and taken $\regA$ to nearly half the boundary, $\varphi_\regA = 1.503$, at $t=0$ (thick red curve). The shaded regions on the boundary cylinder are $\domdA$ and $\domdAc$ respectively. The extremal surface is the thick blue curve, while the purple curves are the rims of the causal wedge (causal information surfaces) for $\regA$ and $\regAc$ respectively. A few representative generators are provided for orientation: the blue null geodesics generate the boundary of the causal wedge for $\regA$ while the green ones do likewise for $\regAc$. The orange generators in the middle of the spacetime generate the boundary of the causal shadow region $\shadow_{\entsurf}$.
}
\label{f:causalshadow}
\end{center}
\end{figure}
The second bulk causal domain which will play a major role in our discussion below is a region we call the {\em causal shadow} $\shadow_{\entsurf}$ associated with the entangling surface $\entsurf$. We define this region as the set of points in the bulk $\bulk$ that are spacelike-related to both $\domdA$ and $\domdAc$, i.e.,
\begin{align}
\shadow_{\entsurf}
& := \left(\bulkJ^+[\domdA]\cup\bulkJ^-[\domdA] \cup \bulkJ^+[\domdAc]\cup\bulkJ^-[\domdAc]\right)^c
\nonumber \\
 &= \bulksplsep[\domdA \cup \domdAc]\,.
\label{shadowdef}
\end{align}	
For a generic region $\regA$ in a generic asymptotically AdS spacetime, the causal shadow is a codimension-zero spacetime region; see Fig.~\ref{f:causalshadow} for an illustrative example.\footnote{ The bulk metric used in the plot for Fig.~\ref{f:causalshadow} is 
\begin{equation*}
ds^2 =\frac{1}{\cos^2\rho} \left(-f(\rho)\, dt^2 + \frac{d\rho^2}{f(\rho)} + \sin^2\rho \, d\varphi^2 \right) ,\qquad f(\rho) = 1-\frac{1}{2}\, \sin^2(2\,\rho)\,.
\end{equation*}	
The matter supporting this geometry satisfies the null energy condition as can be checked explicitly. \label{fn:metric}
 } 
 In certain special (but familiar) situations, such as spherically symmetric regions in pure AdS (where $\rhoA$ is unitarily equivalent to a thermal density matrix), it can degenerate to a codimension-two surface. In such special cases, the entanglement wedge and the causal wedge coincide \cite{Hubeny:2012wa}.
In general, the causal information surface for $\regA$ and that for $\regAc$ comprise the edges of the causal shadow. For a  generic pure state these causal information surfaces each recede from $\extr$ towards their respective boundary region but approach each other near the AdS boundary.  Hence the geometrical structure of $\shadow_{\entsurf}$, described in language of a three-dimensional bulk, is a ``tube'' (connecting the two components of $\entsurf$) with a diamond cross-section, which shrinks to a point where the tube meets the AdS boundary at $\entsurf$.

For topologically trivial deformations of AdS, in the absence of $\extr$ (i.e.\  when the state is pure and $\regA = \Sigma$) the causal shadow disappears, but intriguingly, even when $\regA$ is the entire boundary Cauchy slice, the causal shadow can be nontrivial. This occurs for example in the \AdS{3}-geon spacetimes\footnote{ 
Since these describe pure states, the presence of a causal shadow region does not necessarily guarantee the presence of an extremal surface {\it whose area gives the entanglement entropy} contained within it.  
However, there will be {\it some} extremal surface spanning this region.} \cite{Balasubramanian:2014hda} and in  perturbations of the eternal AdS black hole, such as those studied by \cite{Shenker:2013pqa}. In such  a situation we simply define the casual shadow of the entire boundary (dropping the subscript) as
\begin{align}
\shadow := \bulksplsep[ \bdy] =\left(\bulkJ^+[{\bdy}]\cup\bulkJ^-[\bdy]\right)^c
\label{}
\end{align}	
Here $\bdy$ is understood generally to include multiple disconnected components; the causal shadow is the region spacelike separated from points on all the boundaries.

\subsection{Causality constraints on extremal surfaces}
\label{sec:bulkextc}

Having developed the various causal concepts which we require, let us now ask what the constraints of field-theory causality concerning entanglement entropy translate to in the bulk. The first constraint is that $S_\regA$ should be a wedge observable, i.e.\ if $D[\regA]=D[\regA']$ then $S_\regA=S_{\regA'}$. For this to hold in general, we need $\extr={\cal E}_{\regA'}$. The second concerns perturbations of the field-theory Hamiltonian.
Such perturbations will source perturbations of the bulk fields, including the metric, that will travel causally with respect to the background metric. In particular, disturbances originating in $\domdA$ will be dual to bulk modes propagating in $\bulkJ^+\big[\domdA\big]$ (if we fix the state in the far past) or in $\bulkJ^-\big[\domdA\big]$ (if we fix the state in the far future). If either of these bulk regions intersected $\extr$, the dual of local operator insertions in $D[\regA]$ could change the area of $\extr$, meaning that the HRT proposal would be inconsistent with causality in the QFT. By the same token, the extremal surface cannot intersect  $\bulkJ^+\big[\domdAc\big]$ or  $\bulkJ^-\big[\domdAc\big]$. Since the region complement to union of the causal sets  $\bulkJ^\pm[D[\regA]], \bulkJ^{\pm}[D[\regAc]]$ is the set of points that are spacelike related to $D[\regA]\cup D[\regAc]$, we learn that 
\begin{equation}
\extr \splrel \domdA \;\cup \domdAc \,.
\label{bulkcausal1}
\end{equation}	
In others words, using \eqref{shadowdef} we can say that $\extr$ has to lie in the causal shadow of $\entsurf$
\begin{equation}
\extr \subset \shadow_{\entsurf}\,.
\label{}
\end{equation}	

It is known, based on properties of extremal surfaces, that $\extr$ lies outside the causal wedges $\CWA$ and $\CWAc$ \cite{Hubeny:2012wa,Wall:2012uf, Hubeny:2013gba}. This leaves open the possibility that the surface could still lie in the causal future (or past) of the boundary domain of dependence of $\regA$ or $\regAc$. A particular worry arises in explicit examples in Vaidya-AdS geometries where the extremal surface lies on the boundary of $\bulkJ^+\big[\domdA \big]$. This then leaves open the question whether one might indeed be able to push $\extr$ into a causally forbidden region, by introducing appropriate deformations in $\domdA$.  A theorem of Wall \cite{Wall:2012uf} (Theorem 6 of the reference), guarantees that this does not occur (modulo some assumptions). 

We will prove an essentially equivalent statement in \S\ref{sec:results}, directly for extremal surfaces in an asymptotically AdS spacetime. The main result however can be stated in terms of three simple causal relations:
\begin{equation}
\begin{split}
	\bulkD[\homsurfA] \cap \bdy & =  \domdA \\
	\bulkD[\homsurfAc] \cap \bdy & =  \domdAc \\
	\bulkJ^{\pm}[\extr] \cap \bdy & =  J^{\pm}[\entsurf]\,.
\end{split}	
\label{finalrels}
\end{equation}
In other words, the causal split of the bulk into spacelike- and timelike-separated regions from $\extr$ restricts to the boundary at precisely the boundary split  (\ref{bdy4d}).
Given the decomposition (\ref{bdy4d}), these causal relations imply that perturbations in $D[\regA]\cup D[\regAc]$ are not in causal contact with $\extr$. So, as required, the extremal surface lies in the causal shadow.

As a consequence of this theorem, we will also show that, if there is a spacelike region $\regA' $ such that $D[\regA'] = D[\regA]$, then there is a bulk region ${\cal R}_{\regA'}$ such that $\partial{\cal R}_{\regA'}= \regA'\cup\extr$, so $\extr$ is spacelike-homologous to $\regA'$. Thus, the HRT formula gives the same entanglement entropy for $\regA'$ and $\regA$, as required on the field-theory side.

\subsection{Entanglement for disconnected boundary regions}
\label{sec:gexpt}

A striking consequence of the theorems discussed above emerges when we consider spacetimes with two boundary components, and let $\regA$ be (a Cauchy slice for) all of one component.

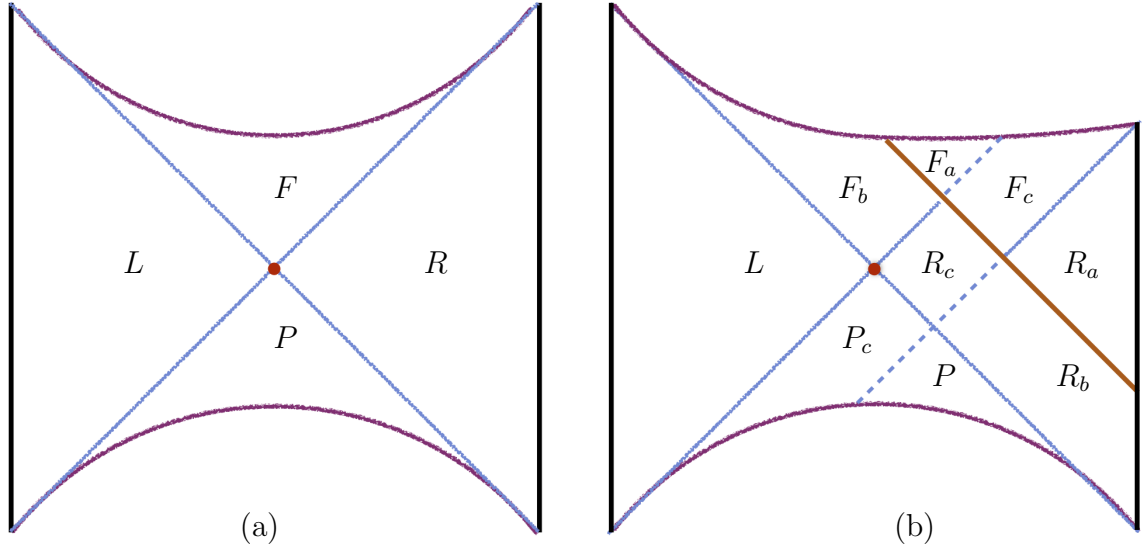
\begin{figure}
\begin{center}
\includegraphics[width=6in]{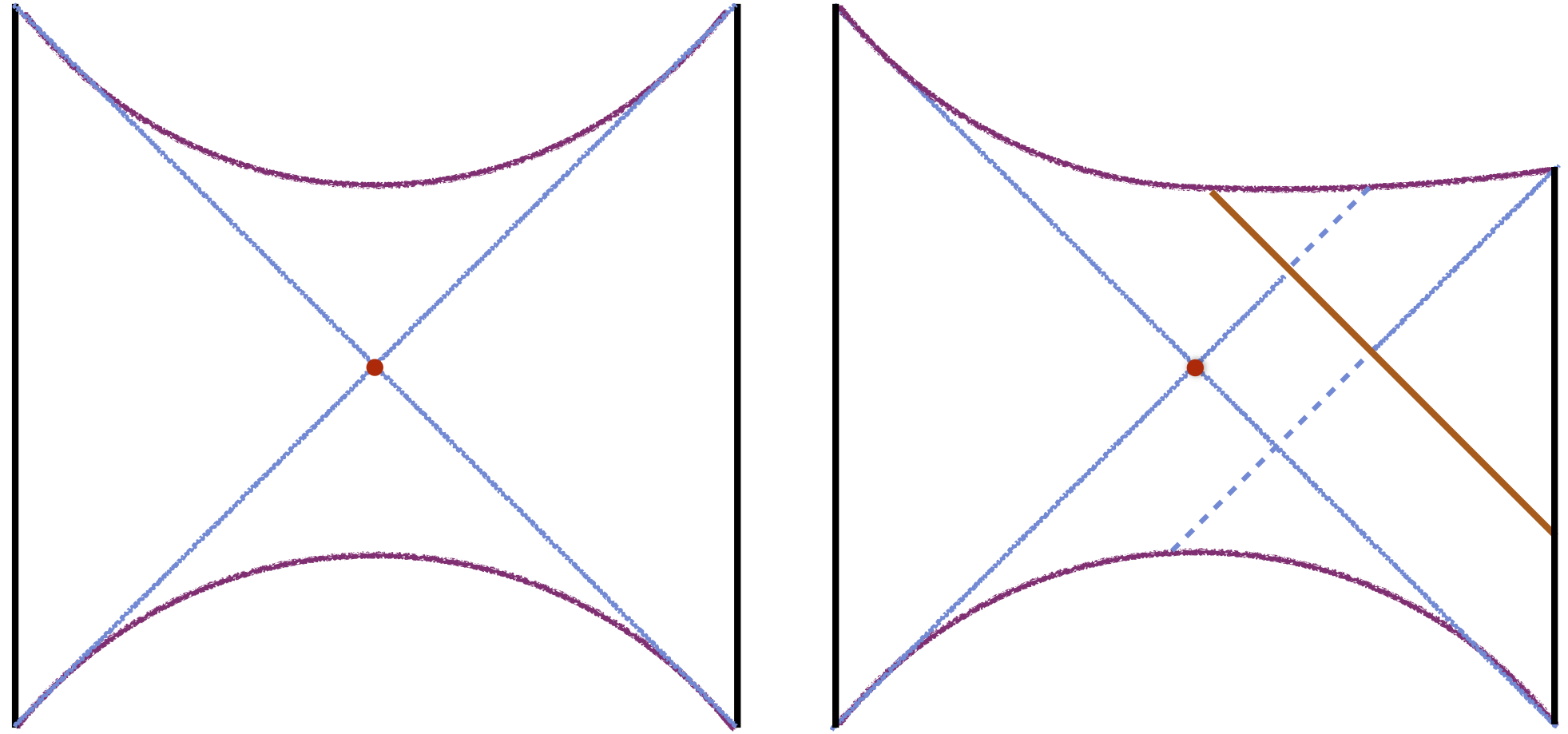}
\begin{picture}(0,0)
\setlength{\unitlength}{1cm}
\put (-12.2,0) {(a)}
\put (-3.5,0) {(b)}
\put(-11.75,4.5){$F$}
\put(-11.75,2.5){$P$}
\put(-9.75,3.5){$R$}
\put(-13.75,3.5){$L$}
\put(-4.25,4.5){$F_b$}
\put(-3.1,4.85){$F_a$}
\put(-2.05,4.5){$F_c$}
\put(-3,2){$P$}
\put(-4.2,2.5){$P_c$}
\put(-1.25,3.5){$R_a$}
\put(-3.15,3.5){$R_c$}
\put(-1.35,2){$R_b$}
\put(-5.5,3.5){$L$}
\end{picture}
\caption{
Sketch of Penrose diagram for {\bf (a)} static eternal \SAdS{} and {\bf (b)}  `thin shell' Vaidya-\SAdS{}, with the various regions labeled.  The AdS boundaries are represented by vertical black lines, the singularities by purple curves, the horizons by diagonal blue lines, and the `shell' in the Vaidya case by diagonal brown line.
}
\label{f:VaidyaSAdSPD}
\end{center}
\end{figure}

As a starting point,  consider the eternal \SAdS{d+1} black hole in the Hartle-Hawking state, with a Penrose diagram shown in  Fig.~\ref{f:VaidyaSAdSPD}(a) below.  The left and right boundaries of the diagram each have the topology ${\bf S}^{d-1}\times \RR$. This  geometry is believed to be dual to the CFT on the product spatial geometry ${\bf S}^{d-1}_L\times {\bf S}^{d-1}_R$, in the entangled ``thermofield double" state \cite{Horowitz:1998xk,Balasubramanian:1998de,CarneirodaCunha:2001jf,Maldacena:2001kr}:
\begin{equation}
\ket{{\rm HH }}_{L,R} = \sum_i\, e^{-\frac{1}{2}\,\beta \, E_i} \; \ket{E_i}_L \, \ket{E_i}_R
\label{hhstate}
\end{equation}	
where $\ket{E_i}_{R,L}$ is the energy eigenstate of the CFT on ${\bf S}^{d-1}_{R,L}$.  

Let $\Sigma_R$ lie on the $t=0$ slice of the right boundary, and consider the reduced density matrix for some region $\regA \subset \Sigma_{R}$. Since this is a static geometry, its entanglement entropy $S_\regA$ is computed by a minimal surface $\extr$ which never penetrates past the bifurcation surface ${\cal X}$ of the black hole \cite{Hubeny:2012ry}.\footnote{ Note that the extremal surface does not come arbitrarily close to the horizon---it either includes a component that wraps the horizon, or stays a finite distance away from it \cite{Hubeny:2013gta}.} If we let $\regA$ be the full Cauchy slice of one of the boundaries, say $\regA = \Sigma_R$, the extremal surface  precisely coincides with the black hole bifurcation surface, as indicated in Fig.\ \ref{f:VaidyaSAdSPD}.  Note that $\extr$ lies on the edge of the causally acceptable region since ${\cal X}$ sits at the boundary of both $\CWA$ and $\CWAc$, and therefore constitutes the entire causal shadow for this special case.  

One might now wonder what happens if we deform the state \eqref{hhstate}. This is not an innocuous question. In time-dependent geometries, the global (teleological) nature of the event horizon implies that extremal surfaces anchored on the boundary {\it can} pass through this horizon \cite{Hubeny:2002dg}. Furthermore, as first explicitly shown in \cite{AbajoArrastia:2010yt}, even apparent horizons do not form a barrier to the extremal surfaces. Hence we see that, a priori, in a state which is a deformation of (\ref{hhstate}), $\extr$ is in danger of entering $\CWAc$.

The theorems we have stated above indicate that this does not happen. The question is, how precisely does the extremal surface $\extr$ avoid doing so?  As a first step to answering this, consider a deformation of the static eternal case localized along a null shell emitted from the right boundary at some time.  The corresponding metric is given by the global Vaidya-SAdS geometry, where both the initial 
(prior to the shell) and final (after the shell) spacetime regions describe a  black hole.  
Fig.~\ref{f:VaidyaSAdSPD}b presents a sketch of the Penrose diagram of such a geometry, contrasted with the standard static eternal \SAdS {} black hole (Fig.~\ref{f:VaidyaSAdSPD}a).  The diagonal brown line represents the shell which is sourced at some time on the right boundary and implodes into the black hole (terminating at the future singularity), and the blue lines represent the various (future and past, left and right) event horizons.  
The solid parts of these lines indicate where these event horizons coincide with apparent horizons (as well as isolated horizons); the dashed parts are parts of the event horizon which are not apparent horizons.  

In such a geometry, let us again consider $\regA = \Sigma_R$. Then our theorems guarantee that the extremal surface must lie on the null sheet separating regions $R_c$ and $P_c$: it is again spacelike-separated from both $D[\Sigma_L]$ and $D[\Sigma_R]$. (In fact, since the spacetime prior to the shell is identical to the eternal static case, the extremal surface remains in the same location as for the static case, namely the bifurcation surface where regions $R_c$ and $L$ touch.) The situation is again marginal, much like the original undeformed case.  Indeed, any perturbation to \SAdS{} which emanates from (or reaches to) the right boundary cannot change the location of the original extremal surface by causality; it could at most generate a new extremal surface.

\begin{figure}
\begin{center}
\includegraphics[width=4in]{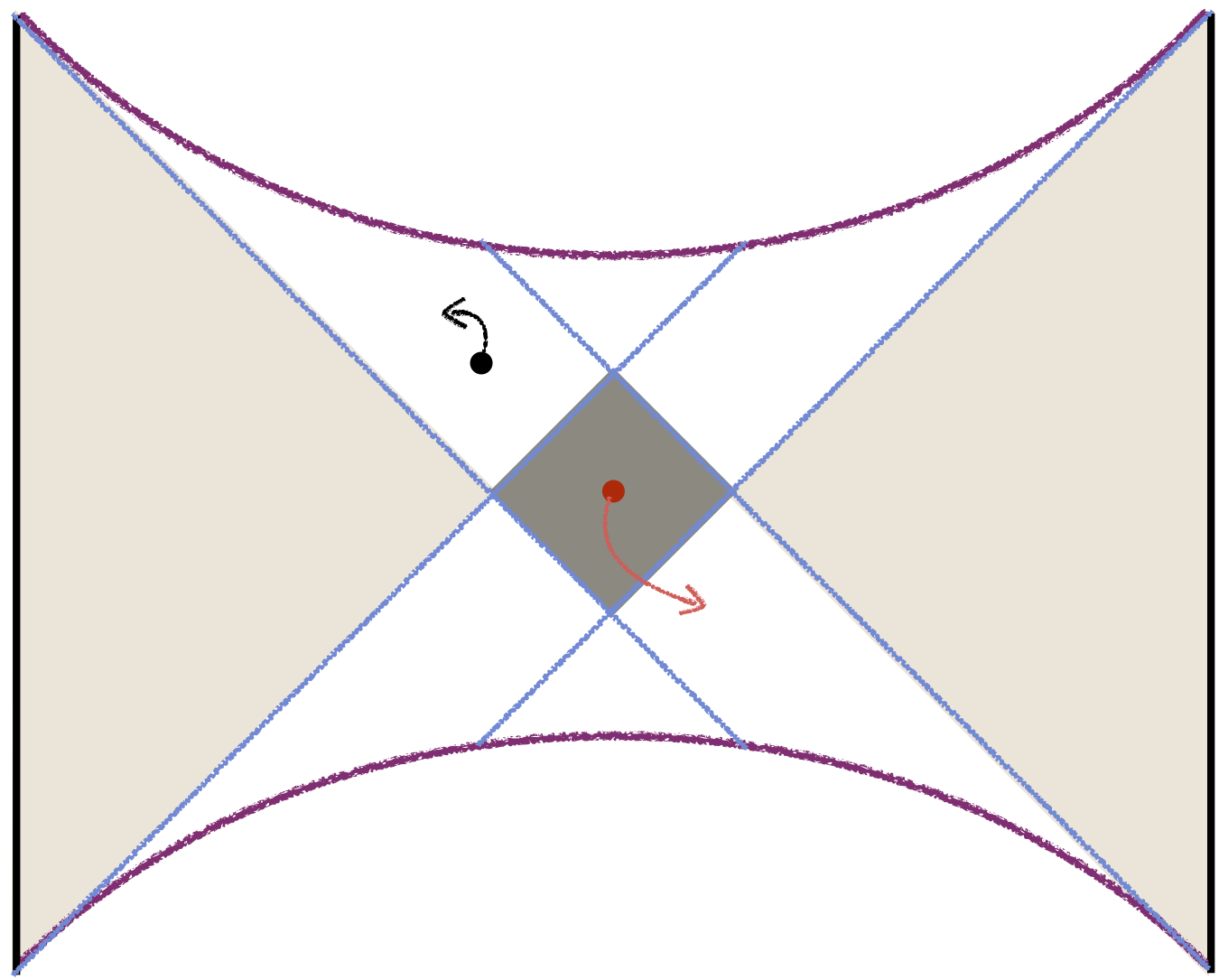}
\setlength{\unitlength}{0.1\columnwidth}
\begin{picture}(0.3,0.4)(0,0)
\put(0.25,3){\makebox(0,0){CFT$_R$}}
\put(-7.15,3){\makebox(0,0){CFT$_L$}}
\put(0.,2){\makebox(0,0){$\regA$}}
\put(-2.65,2){\makebox(0,0){$\color{red}{\extr}$}}
\put(-3.4,2.9){\makebox(0,0){$\shadow$}}
\put(-1.5,2.6){\makebox(0,0){$\CWA$}}
\put(-5.5,2.6){\makebox(0,0){$\CWAc$}}
\put(-4.6,3.6){\makebox(0,0){${\cal F}_\regA$}}
\end{picture}
\caption{
Sketch of Penrose diagram for a symmetric Vaidya-\SAdS{}  geometry obtained by imploding null shells to the past and future from both boundaries. The crucial new feature of note is the presence a {\em causal shadow} region that is spacelike separated from both boundaries. We have also indicated the extremal surface $\extr$ for the region $\regA = \Sigma_R$ in red at the center of the figure and ${\cal F}_\regA$ is a ${\bf S}^{d-1}$ of finite area in the causal future of the left boundary. The lightly shaded regions are the causal wedges associated with $\regA$ and $\regAc$ respectively. 
}
\label{f:VaidyaSAdSSym}
\end{center}
\end{figure}
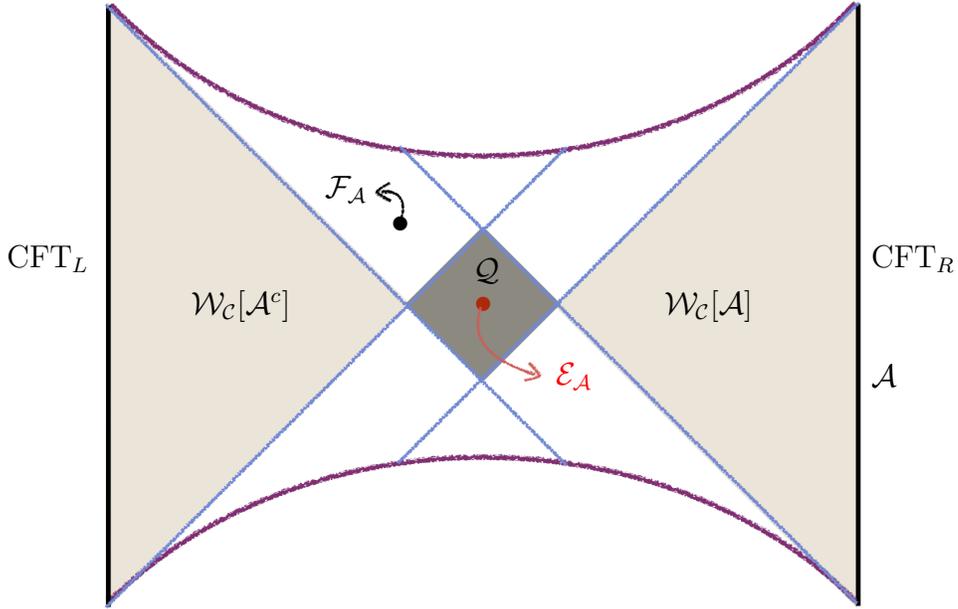

A less marginal case occurs when we symmetrically perturb both copies of the CFT as above.  Consider a perturbation at $t = 0$ such that spherically symmetric null shells are emitted both to the past and future on both sides of the diagram.  One then obtains the Penrose diagram shown in Fig.~\ref{f:VaidyaSAdSSym}; this has time-reflection symmetry about $t=0$, symmetry under exchanging the left and right sides, and the $SO(d)$ 
rotational symmetry.

According to the theorems above, the extremal surface must be spacelike-separated from both boundaries, when we take $\regA = \Sigma_R$. Using both time and space reflection symmetry, it is clear that $\extr$ must sit in the center of the {\em causal shadow} $\shadow$ of the two boundaries, spacelike separated from both.   

In the general case of spherically symmetric spacetime (even in the absence of time or space reflection symmetry) there is an easy proof of our claim that $\extr$ must lie in the causal shadow.
We proceed by contradiction: suppose that a spherical extremal surface $\extr$ lies in $\bulkJ^+\left[\Sigma_L\right]$. This means that 
on a Penrose diagram, it lies somewhere in the top-left region; say it is the surface ${\cal F}_\regA$ indicated in Fig.~\ref{f:VaidyaSAdSSym} (which by rotational symmetry is a copy of ${\bf S}^{d-1}$). Let us then consider the past congruence of null normal geodesics from ${\cal F}_\regA$ towards $\bdy_L$. Since we assume that ${\cal F}_\regA$ candidate surface lies in $\bulkJ^+\left[\Sigma_L\right]$, past-going null congruences from the surface intersect $\bdy_L$ on a spacelike codimension-one surface. In other words, the area of the spheres grows without bound along this past-directed congruence.

However,  by definition, for an extremal surface the initial expansion is vanishing. Moreover, if the matter in the spacetime satisfies the null energy condition,then it also follows that the area along the congruence is guaranteed not to grow. Nor can the area go to zero along the congruence, since the area of the $S^{d-1}$ represented by each point on the Penrose diagram is finite. It therefore follows that our assumption about $\extr$ penetrating $\bulkJ^+\left[\Sigma_L\right]$ must be erroneous; ${\cal F}_\regA$ cannot be an extremal surface. Running a similar argument for the other unshaded regions in Fig.~\ref{f:VaidyaSAdSSym}, we learn that the extremal surface must indeed lie in the causal shadow region, as denoted by the red surface $\extr$.

Indeed, in this particular case, the extremal surface lies at the point on the Penrose diagram where the future and past apparent horizons meet---the ``apparent bifurcation surface''. The fact that it lies in the causal shadow is a consequence of the familiar fact that the apparent horizon can never be outside the event horizon, applied to both future and past horizons.

While the above result relied on the special properties of spherically symmetry (both of the spacetime and the null congruences therein), the theorems we prove in \S\ref{sec:results} will establish this in full generality.

In the next two sections we set out to prove the theorems stated in \S\ref{sec:bulkextc}.  The proof in our spherically symmetric case indicates that understanding null congruences leaving the extremal surface might play a key role.  We will therefore spend some time in \S\ref{sec:examples} examining null congruences emanating from bulk codimension-two surfaces in \AdS{3}, in order to develop a picture of the relevant causal domains, before embarking on a general proof in \S\ref{sec:results}.

\section{Null geodesic congruences in \AdS{3}}
\label{sec:examples}

In this section, we consider null geodesic congruences emanating from curves in \AdS{3} that are anchored at the boundary.  Our aim is to build some intuition about such congruences in a simple setting, since their properties will play a crucial role in the proofs in what follows. Readers familiar with the general statements are invited to skip ahead to the abstract discussion.

We work in the Poincar\'e patch of \AdS{3} with the standard metric:
\begin{equation}
ds^2 = \frac{1}{z^2} \left( -dt^2 + dx^2 + dz^2 \right)
\label{fgads3}
\end{equation}	
Since our aim is to understand specifically the (causal) boundary of  bulk causal domains, we are going to examine properties of null geodesic congruences.  In particular, for a spacelike codimension-one  region $R\subset\bulk$ which is anchored on the AdS boundary, 
the domain of dependence $\bulkD[R]$ is bounded by a family of  outgoing null geodesics emanating from $\partial R$, up to the point where each geodesic encounters a caustic or intersects another generator.\footnote{ 
The latter set of intersections is referred to as cross-over points; the set of these generically form a crossover seam which is codimension-one on this null surface.}

To gain intuition for how these null congruences behave in the context of the extremal surfaces of interest, we examine a more general family of codimension-two surfaces (these are curves in \AdS{3}) which in the above coordinates are given by 
\begin{equation}
x^2 + \frac{z^2}{a^2} = 1 \ , \qquad  t=0
\label{initsurfellipse}
\end{equation}	
parameterized by $a$.  Note that all of these are anchored on the boundary ${\mathbb R}^{1,1}$ at the ends of the interval $\regA = \{(t,x) \in {\mathbb R}^{1,1} \;\mid\; t=0,\; x\in [-1,1]\} $. (For orientation, see the bottom set of curves in Fig.~\ref{f:initsurfs}.) When $a=1$, the surface is a semi-circle, which is simultaneously the causal information surface $\CIS$ defined in \cite{Hubeny:2012wa}, and the extremal surface $\extr$ for the region $\regA$ under consideration.  Surfaces with $a< 1$ lie inside the causal wedge $\CWA$, while those with $a>1$ lie outside i.e., they are spacelike related to $\domdA$. We wish to study the family of null congruences leaving these surfaces, as we vary $a$. The geodesics will be labelled by their starting position $x_0$  and parameterized by an affine parameter $\lambda$ (fixed such that we have unit energy along each geodesic). 

\subsection{Explicit solutions for geodesic congruences}

Since the $a=1$ surface is extremal, the null expansion $\Theta(\lambda; a=1) = 0$ for each generator.  For the surfaces with $a < 1$, closer to the boundary, we expect that the expansion is positive and the congruence intersects the boundary in a spacelike curve inside 
$\domdA = \{(t,x) \in {\mathbb R}^{1,1} \mid |t \pm x| \leq 1 \}$.   
For curves with $a>1$, long ellipse, we expect the expansion to be negative.  The resulting congruence should develop a caustic before reaching the boundary.

\begin{figure}[ht!]
\begin{center}
\includegraphics[width=2.in]{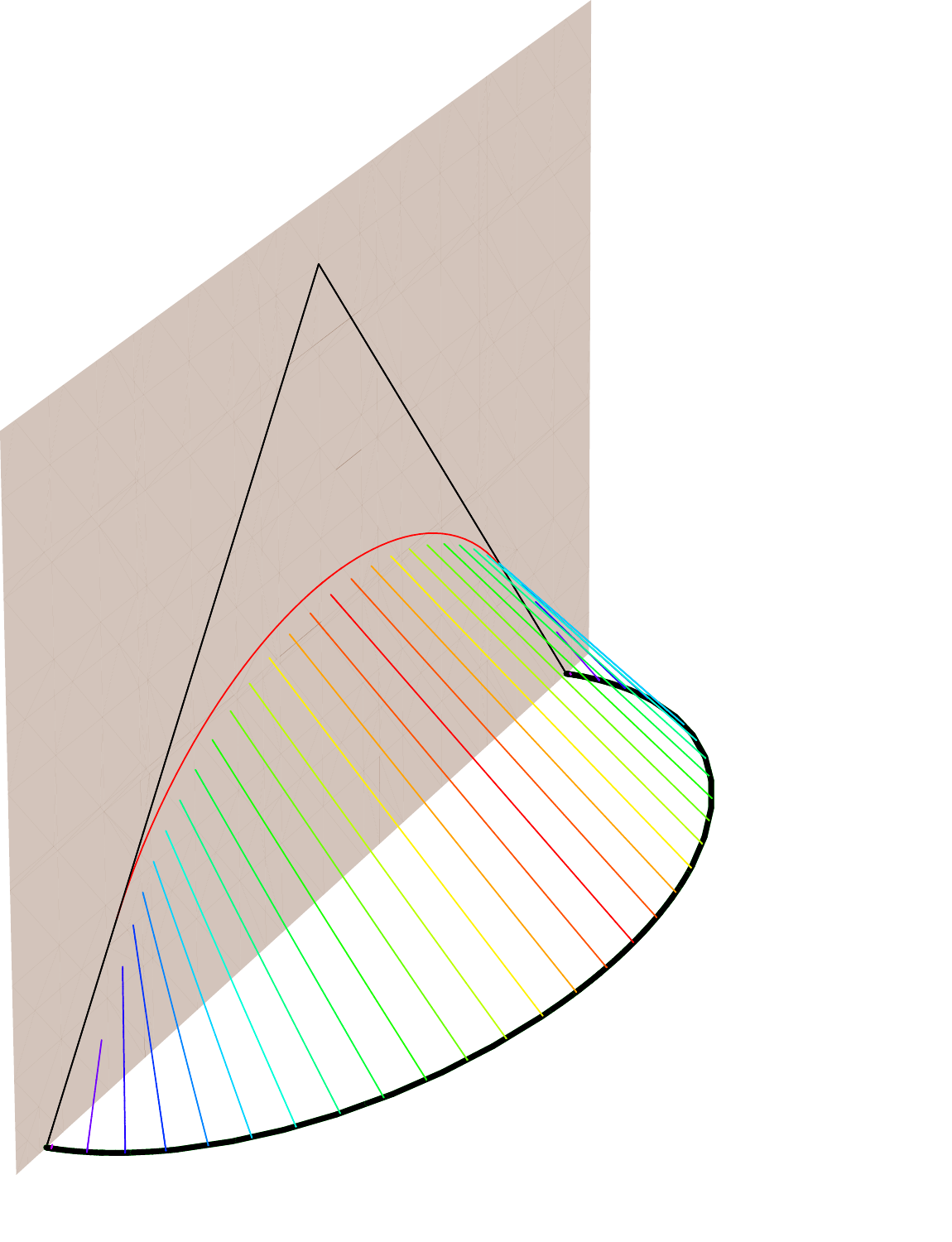}
\hspace{2cm}
\includegraphics[width=2in]{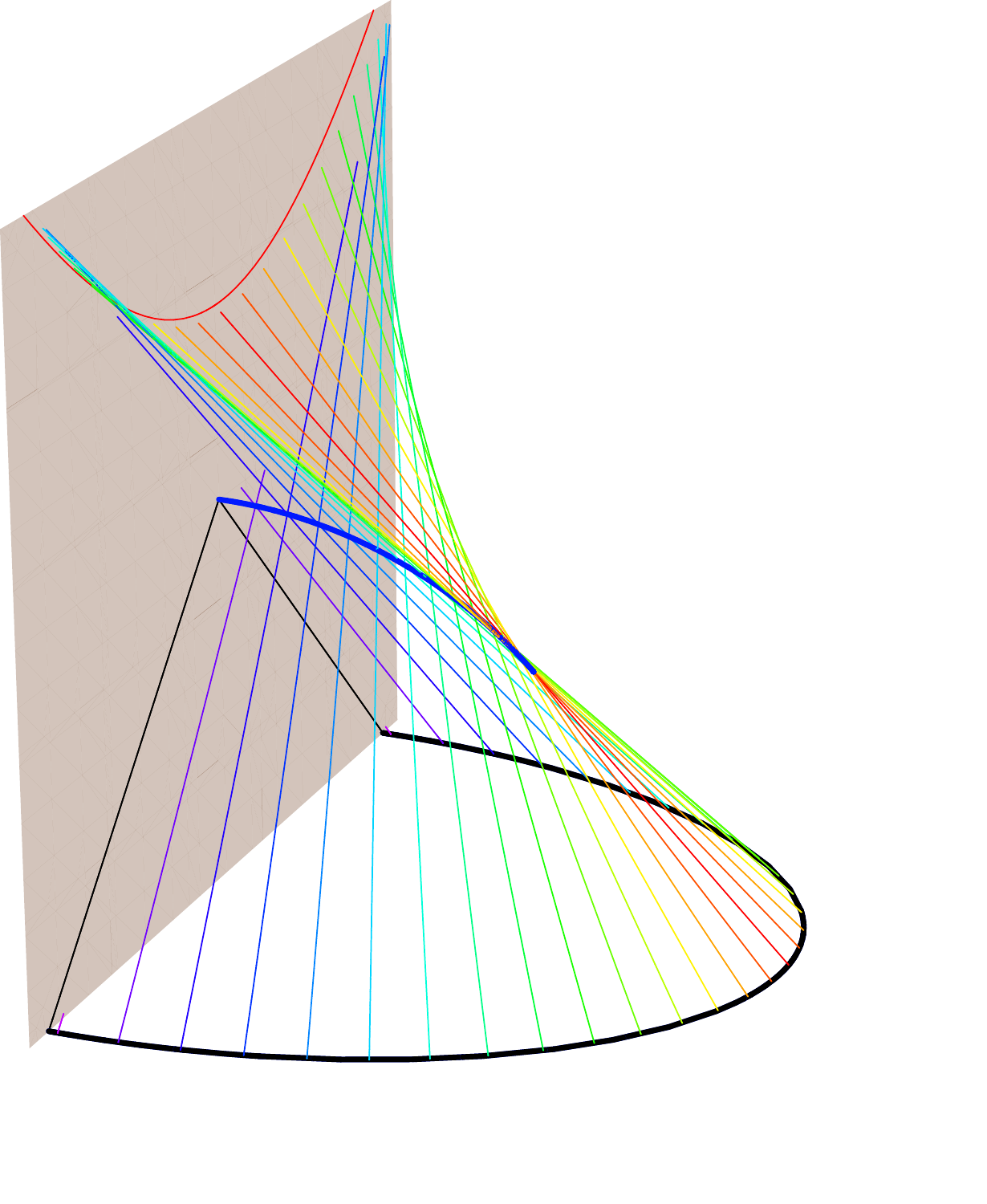}
\caption{
Null normal congruence from the initial surface given by \eqref{initsurfellipse} with $a=0.5$ (left) and $a=1.5$ (right).  The initial surface is the bold black curve on the bottom, the boundary is the shaded plane on the left in each plot (with the domain of dependence $D^+[\regA]$ boundary indicated by the thin black lines), the individual geodesics are the thin lines color-coded by $x_0$, their endpoints on the boundary are depicted by the red curve, and finally the seam of crossover points where generators intersect for $a>1$ is the blue thick curve.  (The generators are cut off at a finite value of $\lambda \approx 64$, so in the plot they don't look like they reach all the way to the boundary.)
}
\label{f:congruence}
\end{center}
\end{figure}

Due to the relative simplicity of the set-up, we can confirm these expectations explicitly.  Since everything is time-symmetric, let us consider just the future-directed outgoing congruence:
\begin{equation}\begin{split}
z(\lambda) &= \frac{a \, \sqrt{1-x_0^2} \, \sqrt{1-x_0^2+a^2 \, x_0^2}}
{a \, (1-x_0^2) \, \lambda +\sqrt{1-x_0^2+a^2 \, x_0^2}} 
\\
x(\lambda) &= x_0 \, \frac{a \, (1-a^2) \, (1-x_0^2) \, \lambda + \sqrt{1-x_0^2+a^2 \, x_0^2}}
{a \, (1-x_0^2) \, \lambda +\sqrt{1-x_0^2+a^2 \, x_0^2}}  
\\
t(\lambda) &= \frac{a^2 \, (1-x_0^2) \,  \sqrt{1-x_0^2+a^2 \, x_0^2} \, \lambda}
{a \, (1-x_0^2) \, \lambda +\sqrt{1-x_0^2+a^2 \, x_0^2}} 
\label{ellipsegeods}
\end{split}\end{equation}	
Note that the endpoints of these generators at $\lambda = \infty$ are given by 
\begin{equation}
z_\infty = 0 \ , \qquad
x_\infty = x_0 \, (1-a^2) \ , \qquad
t_\infty = a \, \sqrt{1-x_0^2+a^2 \, x_0^2}
\label{}
\end{equation}	
A representative plot of the generators is given in Fig.~\ref{f:congruence} for $a=0.5$ (left) and $a=1.5$ (right).  We see that when $a<1$, the generators don't intersect each other before reaching the boundary, and they reach within $D^+[\regA]$.  On the other hand, when $a>1$, the generators intersect in a seam (drawn as thick blue curve, whose explicit expression is given below in \eqref{intersectioncoords}), before reaching the boundary (with the geodesic endpoints indicated by the red curves in Fig.~\ref{f:congruence}). We call the points on this seam the {\em cross-over points}; non-neighbouring geodesics intersect at these points.  This seam terminates in a {\em caustic}, which as always  refers to the locus  where neighbouring geodesics intersect.
%

\subsection{Intersections within congruences}

We can determine the intersection between distinct geodesics in the bulk using the explicit expressions from \eqref{ellipsegeods}.  By symmetry of the set-up, we know that geodesics with opposite values of $x_0$ necessarily intersect, and they must do so at $x=x_\times = 0$.  Solving for the intersection of the pair of geodesics starting from $x_0$ and $-x_0$ we find that they meet at:
\begin{equation}
t_\times = \frac{ \sqrt{1-x_0^2+a^2 \, x_0^2}}{a} \ , \quad
z_\times = \frac{a^2-1}{a} \,  \sqrt{1-x_0^2} \ , \quad
\lambda_\times = \frac{ \sqrt{1-x_0^2+a^2 \, x_0^2}}{a \, (a^2-1) \, (1-x_0^2)}
\label{intersectioncoords}
\end{equation}	
%
\begin{figure}
\begin{center}
\includegraphics[width=2.8in]{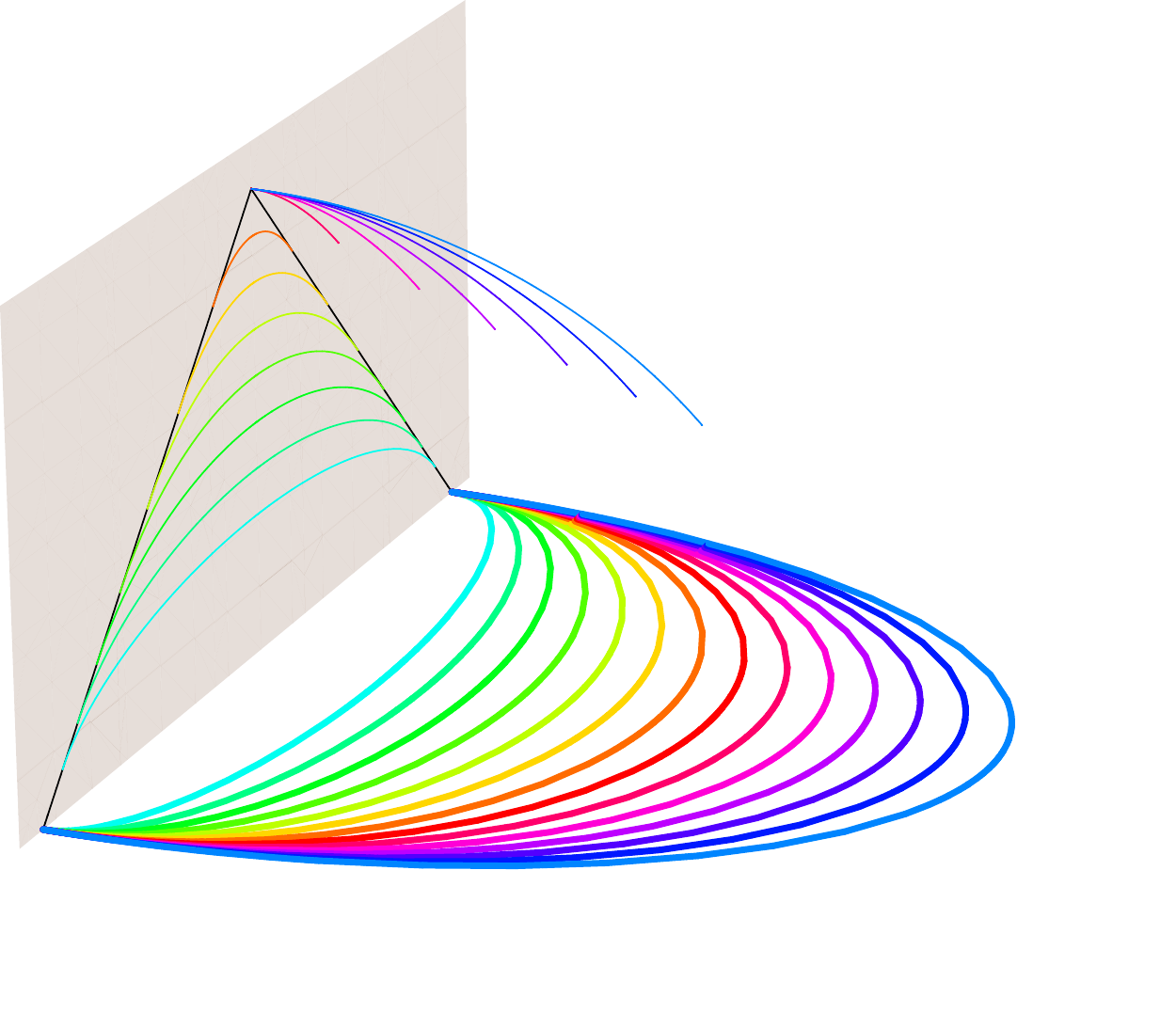}
\caption{
Initial surfaces (thick curves at the bottom, color-coded by $a$), along with endpoints of the generators of the corresponding null congruence:  for $a=1$ (initial surface is the red semi-circle), all generators meet at the tip.  Increasing $a>1$ (color shift towards purple and blue) makes the generators intersect at the seam of cross-over points before reaching the boundary.  On the other hand, decreasing $a<1$ (color shift towards orange and green) makes the generators reach the boundary within $D^+[\regA]$ (depicted as in Fig.~\ref{f:congruence}).
}
\label{f:initsurfs}
\end{center}
\end{figure}
This generates the seam of cross-over points depicted in the right panel of 
Fig.~\ref{f:congruence}, and plotted for various values of $a$ in Fig.~\ref{f:initsurfs} (the top set of curves, color-coded by $a$ corresponding to the initial surface indicated by the thick horizontal curve of the same color).
It is easy to see from \eqref{intersectioncoords} that the cross-over points terminate on the boundary at the future tip of $D^+[\regA]$, i.e., at $z=0, \ x=0, \ t=1$, corresponding to the intersection of the boundary geodesics $x_0=\pm1$.  On the other hand, the cross-over seams for different $a$ start at the point in the bulk when neighbouring geodesics from $x_0 \simeq 0$ intersect which happens at\
\begin{equation}
x_\times = 0 \ , \qquad
t_\times = \frac{1}{a} \ , \qquad
z_\times = \frac{a^2-1}{a} \ , \qquad
\lambda_\times = \frac{1}{a \, (a^2-1)}
\label{causticbottom}
\end{equation}	

To summarize, depending on whether $a$ is greater or less than 1, the congruence has qualitatively different behaviour, as illustrated in Fig.~\ref{f:initsurfs}.  For $a<1$ (depicted by colors from red toward green), the congruence reaches the boundary inside $D^+[{\cal A}]$, while for $a>1$, the generators intersect each other at the seam of crossover points (depicted by colors from red toward purple).  At precisely $a=1$, all generators reach the boundary at the future tip of $D^+[{\cal A}]$, namely $z=0, \ x=0, \ t=1$.

\subsection{Expansion of congruences and caustics}

Let us now analyze the expansion along this congruence. This can be calculated as the change in area along the wavefront
\begin{equation}
\Theta(\lambda,x_0) = \frac{1}{A(\lambda,x_0)} \, \frac{\partial}{\partial \lambda}A(\lambda,x_0)
\label{ellipseThetadef}
\end{equation}	
with
\begin{equation}
A(\lambda,x_0) = \int_{x_0}^{x_0+\delta x}
\sqrt{\frac{-t'(\lambda ,{\tilde x}_0)^2+x'(\lambda ,{\tilde x}_0)^2+z'(\lambda ,{\tilde x}_0)^2}{z^2(\lambda ,{\tilde x}_0) }} \, d{\tilde x}_0
\label{}
\end{equation}	
where $t'(\lambda,x_0) \equiv \frac{\partial}{\partial x_0} t(\lambda;x_0)$ etc., using the expressions given in \eqref{ellipsegeods}. While one can numerically solve for $\Theta(\lambda)$ it is easier to obtain the solution for small $\lambda$ and evolve using the Raychaudhuri equation. 

\begin{figure}
\begin{center}
\includegraphics[width=2.5in]{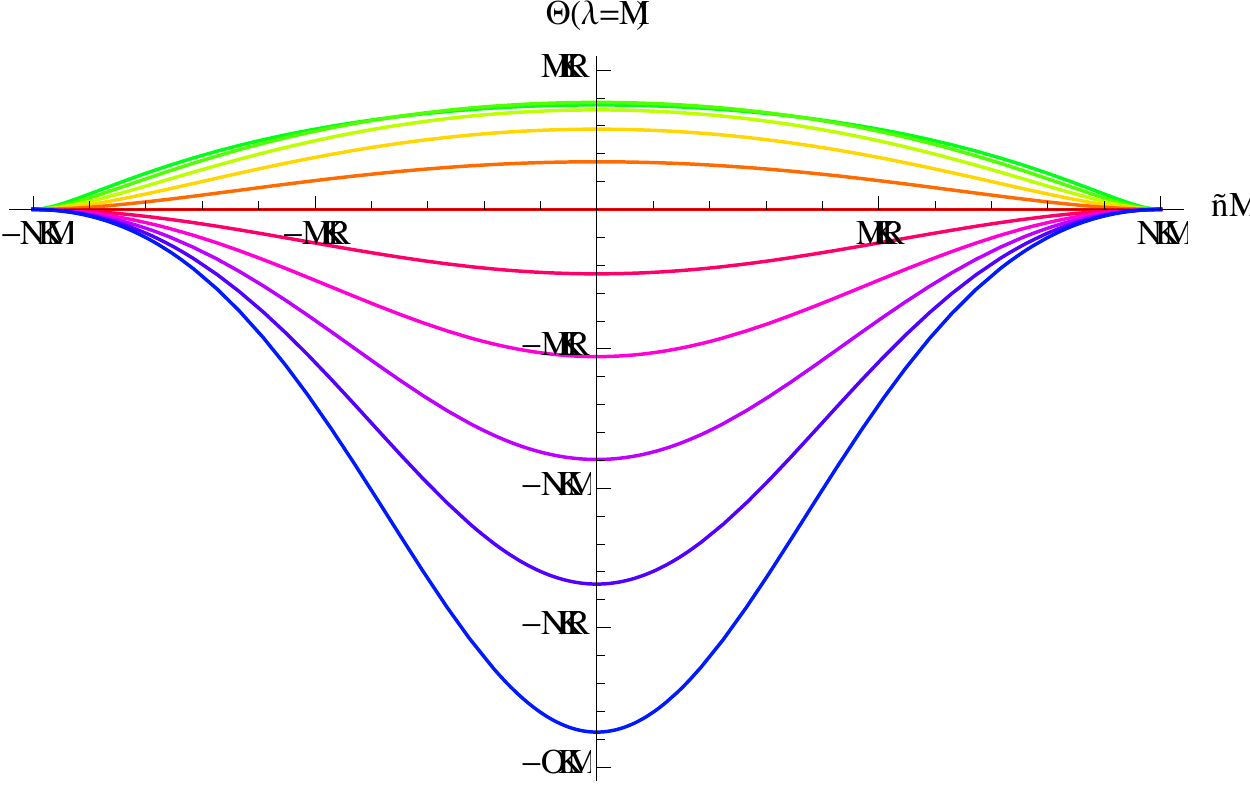}
\hspace{1cm}
\includegraphics[width=2.5in]{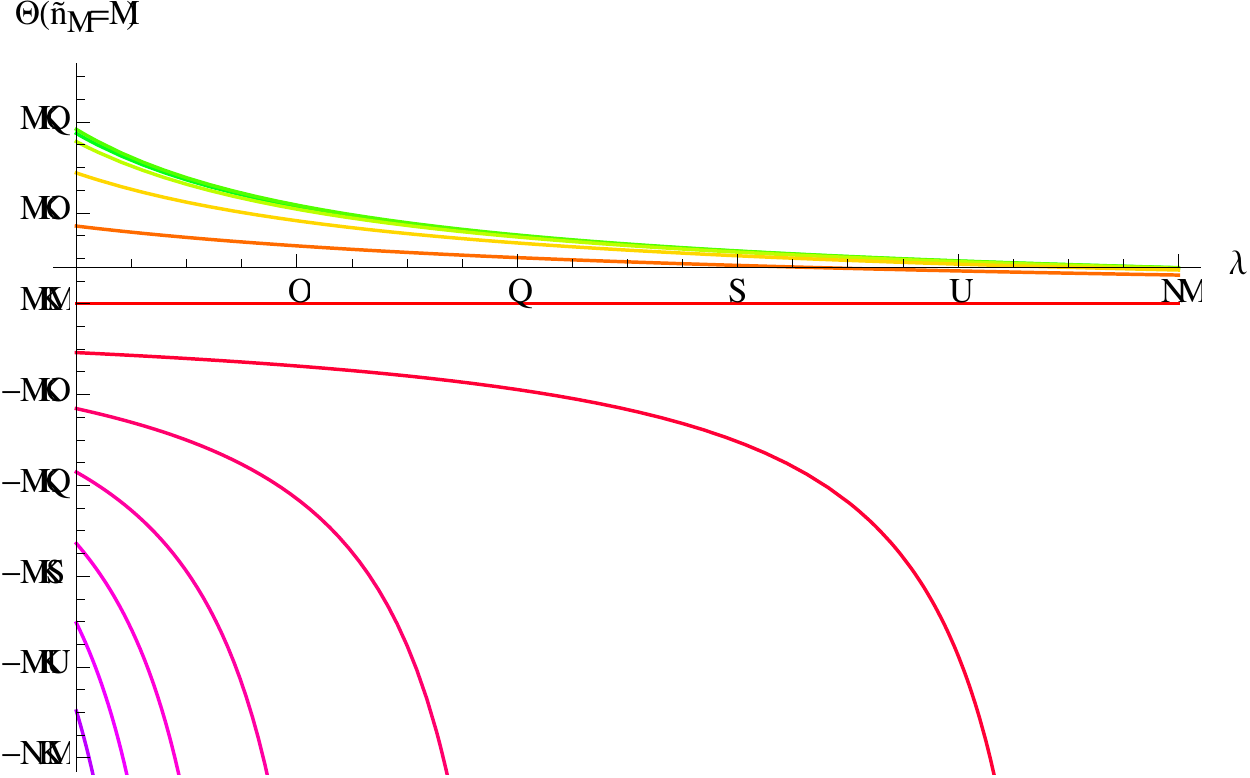}
\caption{
Expansion $\Theta(\lambda; x_0)$ along the generators for various values of $a$ (color-coded by $a$ as in Fig.~\ref{f:initsurfs}).
On left, we show the expansion from the initial surface $\lambda =0$ as a function of the starting position $x_0$.  On right, we fix $x_0=0$ as plot the evolution of $\Theta(\lambda)$ along the radial generator.
}
\label{f:expansion}
\end{center}
\end{figure}

Near $\lambda=0$, the leading order expression for $\Theta$ is: 
\begin{equation}
\Theta_0 \equiv \Theta(\lambda=0) = 
\frac{a \, (1-a^2) \, (1-x_0^2)^2}{(1-x_0^2+a^2 \, x_0^2)^{3/2}}
\label{initexp}
\end{equation}	
This is plotted in the left panel of Fig.~\ref{f:expansion} (with same color-coding by $a$ as employed in Fig.~\ref{f:initsurfs}). 
At the ends of the interval $x_0=\pm 1$, $\Theta_0$ vanishes (which is to be expected since the congruence approximates a larger one with $a=1$), while $\Theta_0$ reaches its extremum at the midpoint, $x_0=0$ (again, expected by symmetry), where  $\Theta_0(x_0=0) = a \, (1-a^2)$. Furthermore, $\Theta_0$ is positive for $a<1$ and negative for $a>1$; that is, the congruences are expanding for $a<1$ and converging for $a>1$). The former make it out to the boundary without intersecting, while the latter have a seam of cross-overs. As we will see below, the geodesics end in a curve of caustics, which touches the seam of cross-overs at the endpoint of the latter.

Given $\Theta_0$ as our initial condition, it is straightforward to solve the Raychaudhuri equation
\begin{equation}
\frac{d\Theta}{d\lambda} = -\Theta^2 -2\,\sigma_{ab}\, \sigma^{ab} - R_{ab} \,\xi^a\, \xi^b
\label{rceqn}
\end{equation}	
to find the expansion along the geodesics. Here $\xi^a$ is the tangent vector to the null geodesics and $\sigma_{\mu\nu} $ is the shear of the congruence. For a one-dimensional congruence the shear  trivially vanishes and the Ricci tensor contracted with null tangents likewise vanishes upon using the bulk equations of motion  $R_{ab} = -2 \, g_{ab}$, so (\ref{rceqn}) simplifies to:
\begin{equation}
\frac{d \Theta}{d\lambda} = - \Theta^2 
\qquad \Rightarrow \qquad
\Theta(\lambda) = \frac{\Theta_0}{1+ \Theta_0 \, \lambda}
\label{}
\end{equation}	
Using \eqref{initexp}, we find:
\begin{equation}
\Theta(\lambda, x_0) = 
\frac{a \, (1-a^2) \, (1-x_0^2)^2}
{(1-x_0^2+a^2 \, x_0^2)^{3/2} + a \, (1-a^2) \, (1-x_0^2)^2 \, \lambda}
\label{ThetaEllipse}
\end{equation}	
In Fig.~\ref{f:expansion} we have plotted this as a function of $\lambda$ for $x_0 = 0$, at which  $\Theta = \frac{a \, (1-a^2)}{1+ a \, (1-a^2) \, \lambda}$.

For $a > 1$, we expect the congruence to develop a caustic where the expansion diverges. This occurs when infinitesimally nearby geodesics intersect each other. Eq. \eqref{ThetaEllipse} shows that this can only occur for $a>1$, where the second term in the denominator is negative for positive $\lambda$. In this case $\Theta(\lambda) \to -\infty$ at a finite value of $\lambda = \lambda_c$, 
\begin{equation}
\lambda_c = \frac{ (1-x_0^2+a^2 \, x_0^2)^{3/2}}{a \, (a^2-1) \, (1-x_0^2)}
\label{}
\end{equation}	
for any $x_0$.  The spacetime coordinates for the points along the congruence where this happens are given by 
\begin{equation}
x_c = (1-a^2) \, x_0^3 \ , \qquad
t_c = \frac{ (1-x_0^2+a^2 \, x_0^2)^{3/2}}{a} \ , \qquad
z_c = \frac{a^2-1}{a} \,  (1-x_0^2)^{3/2} 
\label{divergingTheta}
\end{equation}	
Viewed as a pair of parametric curves parametrized by $x_0$ which starts at $x_0=0$ and ends at $x_0=\pm 1$, the caustic seams are null curves, starting at the intersection point \eqref{causticbottom} and ending on the boundary at $z_c =0$, $x_c = \pm (1-a^2)$, and $t_c = a^2$.  Note that this is a finite distance on the boundary.

The divergence $\Theta \to -\infty$ signifies the presence of conjugate points, but their geometric meaning is a bit obscure in our discussion so far. The reason is as follows:  as we see  in Fig.~\ref{f:congruence} and can check explicitly, we generically have caustics in the neighbourhood of $x_0 \simeq 0$, but more generally encounter  cross-over points from the intersection geodesics symmetrically placed about $x_0 = 0$. The expansion is {\it finite} along the cross-over seam \eqref{intersectioncoords} for $x_0 \neq 0$.  This can be understood by realizing that the expansion is a local property of the nearby geodesics which doesn't know about any other piece of the congruence. So nothing special ought to happen at the cross-over points which are non-local in the congruence, and indeed these are not conjugate points.

\begin{figure}[t]
\begin{center}
\includegraphics[width=2.5in]{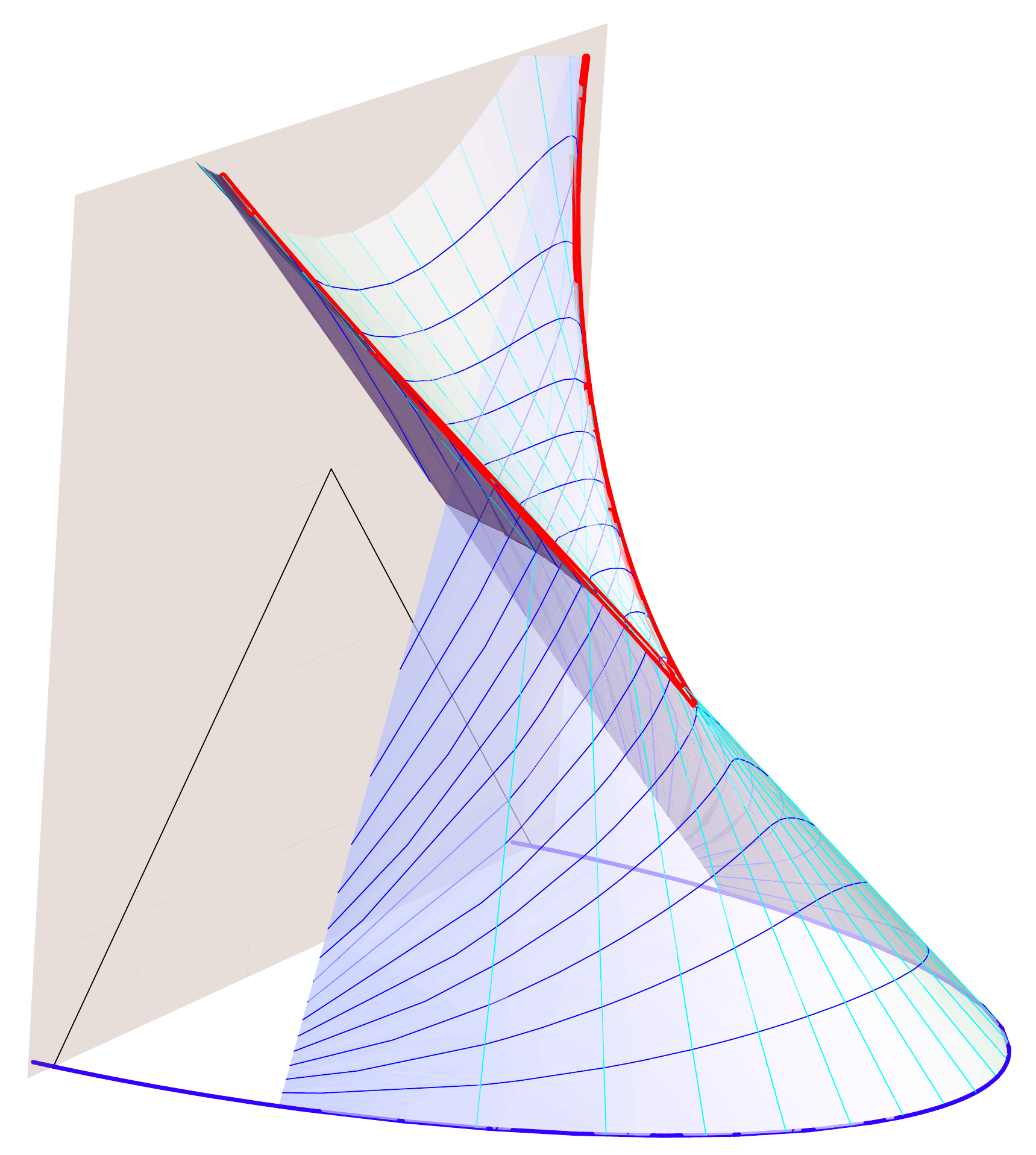}
\caption{
Surface generated by the null normal congruence, along with the locus of points on this surface where the expansion diverges, indicated by the thick red curves.  The cyan contours represent the geodesic generators, while the blue contours are the constant-$\lambda$ wavefronts (we cut off the surface at $|x_0| < 1$ for convenience).
}
\label{f:InftExpansion}
\end{center}
\end{figure}

The clue as to the geometric meaning of $\Theta \to -\infty$ comes from plotting this locus on the surface of the null congruence (continued through the cross-over seam).  This is presented in Fig.~\ref{f:InftExpansion} by the thick red curves.  We see that the surface intersects itself at the cross-over seam, beyond which the constant-$\lambda$ wavefronts form closed loops.  On the sharp flank, these wavefronts turn around and locally become null; this is precisely where $A(\lambda,x_0)$ vanishes and therefore $\Theta \to -\infty$.  

\subsection{Summary}\label{p:ellipsesummary}

The upshot of our calculations can be summarized as follows. Consider the null geodesic congruence emanating from a codimension-two spacelike surfaces ${\cal F}_\regA \subset \bulk$  anchored on the boundary  of a region $\regA$ with $\partial \regA = {\cal F}_\regA \cap \bdy$.
\begin{itemize}
\item If ${\cal F}_\regA \subset \CWA$ then the congruence terminates inside 
$\domdA$ along a spacelike boundary codimension-one surface. 
\item If ${\cal F}_\regA$ lies on the boundary of the causal wedge $\CWA$ then the congruence intersects the boundary on the null surface $\partial \domdA$.
\item If ${\cal F}_\regA \subset \bulksplsep\left[\domdA\right]$ then the congruence finds itself terminated by a seam of cross-over points (and if continued further, would encounter caustic points prior to reaching the AdS boundary). The seam itself however reaches out to the boundary and ends on the future tip\footnote{ 
In higher-dimensional setting, $\domdA$ itself may terminate in a crossover seam rather than a single point, which occurs when the null generators of $\partial \domdA$ on the boundary themselves cross over.
} of $\partial \domdA$.
\end{itemize}
This gives a clear picture of the causal domains for regions bounded by curves inside and outside of $\CWA$.  As we will see in our explicit proof, the extremal surface will in general lie outside of $\CWA$; in special cases it can at best lie on the boundary, but never in the interior, of the causal wedge.

\section{Theorem and proof}
\label{sec:results}

We now get to the main part of the paper where we prove that the extremal surface $\extr$
satisfies the causality requirements discussed in \S\ref{sec:bulkextc}. Our main goal will be to establish the causal relations quoted there in \eqref{finalrels}.  
These will establish for us the consistency of the HRT proposal for computing holographic entanglement entropy. 

In \S\ref{sec:psetup}, we remind the reader of the holographic set-up and of our assumptions. In \S\ref{sec:pnullcong}, we study null geodesic congruences in the bulk and their intersections with the boundary. In particular, since a geodesic that reaches the boundary travels an infinite affine parameter, a non-expanding congruence that reaches the boundary without hitting a caustic must have vanishing shear, and therefore must intersect the boundary at a single point. This allows us to show, using the null energy condition, that the intersection with the boundary of the causal future of an extremal bulk surface equals the causal future of its intersection with the boundary. As a warm-up, we prove a version of the Gao-Wald theorem \cite{Gao:2000ga}. Finally, in \S\ref{sec:pspatial}, we carefully define what we mean by a region and by the spacelike homology condition. We prove that a region $\regA$ implies a natural decomposition of the spacetime into four regions $D[\regA]$, $D[\regAc]$, and $J^\pm[\partial\regA]$. Then, given the spacelike homology condition, and using the results of \S\ref{sec:pnullcong}, we establish the compatibility of the boundary and bulk decompositions, \eqref{finalrels}, and prove that the extremal surface is a wedge observable.

\subsection{Holographic setup}
\label{sec:psetup}

In this subsection we will describe our holographic setup and assumptions.\footnote{ We largely follow the setup and assumptions of section 3 of \cite{Gao:2000ga}, with two exceptions: we remove the null generic condition and we add the condition that the boundary is totally geodesic for null geodesics (assumption (iii) below).}

Let $(\bulk,g_{ab})$ be a connected spacetime, of dimension greater than or equal to 3, that can be embedded in a spacetime $(\bulkC,\tilde g_{ab})$, such that the boundary $\bdy$ of $\bulk$ in $\bulkC$ is a smooth timelike hypersurface in $\bulkC$, and such that $\tilde g_{ab}=\Omega^2 g_{ab}$, where $\Omega$ is a smooth function on $\bulkC$ that vanishes on $\bdy$. (We do not assume that $\bdy$ is connected.) We define $\overM:=\bulk\cup\bdy$. On $\overM$ we have a causal structure induced from $\tilde g_{ab}$, which in $\bulk$ agrees with that induced from $g_{ab}$. We make the following assumptions: 
\begin{enumerate}
\item[(i)] $(\bulk,g_{ab})$ obeys the null energy condition. 
\item[(ii)] $\overM$ is globally hyperbolic.
\item[(iii)] Every null geodesic in $(\bdy,\tilde g_{ab})$ is a geodesic in $(\overM,\tilde g_{ab})$.\footnotemark
\end{enumerate}
 \footnotetext{Assumption (iii) is equivalent to the following property of the extrinsic curvature $K_{ab}$ of $\bdy$ in $\overM$: for any point $p\in\bdy$ and any null vector $k^a$ in the tangent space to $\bdy$ at $p$, $K_{ab}k^ak^b=0$. That it holds for an asymptotically AdS spacetime can be seen by working in Fefferman-Graham coordinates. If we set $\Omega=1/z$, where $z$ is the standard radial coordinate, then $K_{ab}=0$ (so all geodesics in $\bdy$ are geodesics in $\overM$, i.e.\ $\bdy$ is totally geodesic). The property $K_{ab}=0$ is not preserved by Weyl transformations, and so does not hold for a general choice of $\Omega$, but the weaker condition $K_{ab}k^ak^b=0$ does (as can be seen either from a direct calculation or from the fact that the set of null geodesics is invariant under Weyl transformations).}

We begin by showing that $\bdy$ is globally hyperbolic. We omit the proofs, which are very simple, cf., \cite{Galloway:1999bp}. (For brevity, we will only indicate one time direction for each statement below, but the time-reversed statements are clearly equally valid.)

\begin{lemma}\label{domaincontained}For any set $ \Upsilon \subset\overM$, 
$\bulkD^+[\Upsilon] \cap\bdy\subset \domd{\Upsilon\cap\bdy}$.\end{lemma}

\begin{lemma}If $\tilde\Sigma\subset\overM$ is closed and acausal, then $\tilde\Sigma\cap\bdy$ is closed and acausal in $\bdy$.\end{lemma}

\begin{corollary}\label{boundaryCauchy} If $\tilde\Sigma$ is a Cauchy slice\footnote{ We remind the reader that, as explained in footnote \ref{Cauchydef}, throughout this paper we require all Cauchy slices to be acausal, not just achronal.} for $\overM$, then $\tilde\Sigma\cap\bdy$ is a Cauchy slice for $\bdy$.\end{corollary}

\begin{corollary}$\bdy$ is globally hyperbolic.\label{MdotGH}\end{corollary}

\subsection{Congruences of null geodesics}
\label{sec:pnullcong}

In this subsection, we will study null geodesics in $\overM$. Assumption (iii) has the following useful implication:
\begin{lemma}\label{nullgeo}Any null geodesic in $\overM$ either (1) lies entirely in $\bdy$, or (2) does not intersect $\bdy$ except possibly at its endpoints, where it is not tangent to $\bdy$.\end{lemma}

\emph{Proof:} Given a point $p$ in $\bdy$ and a non-zero null vector in the tangent space to $\bdy$ at $p$, there exists a null geodesic in $\bdy$ passing through $p$ with that tangent vector. By assumption (iii), it is a geodesic in $\overM$, and by the uniqueness of geodesics it is the only one. Therefore no null geodesic passing through $\bulk$ can intersect $\bdy$ tangentially. Finally, since $\bdy$ is the boundary of $\overM$ and is smooth, any smooth curve that intersects $\bdy$ at some point without ending there must be tangent to it. $\Box$\bigskip

Now we constrain the behavior of congruences of null geodesics that pass through $\bulk$, using the fact that the metric $g_{ab}$ obeys the null energy condition and the fact that a geodesic that reaches $\bdy$ travels an infinite affine parameter.

\begin{lemma}\label{point}Consider a codimension-one congrence of future-directed null geodesics in $\overM$, each of which lies entirely in $\bulk$ except possibly at its endpoints. Suppose that the part of the congruence in $\bulk$ has the following properties: (1) its expansion with respect to the metric $g_{ab}$ is nowhere positive; (2) at each point, every deviation vector is spacelike and orthogonal to the tangent vector. Then the congruence intersects $\bdy$ on a set of isolated points.\end{lemma}

\emph{Proof:} We begin by working in the metric $g_{ab}$. Since the deviation vectors are everywhere spacelike, the expansion $\Theta$ is finite everywhere. On any geodesic that reaches $\bdy$, the affine parameter goes to infinity, so, by the null energy condition, 
$\Theta$ is nowhere negative, and therefore vanishes everywhere. Again using the null energy condition, the shear therefore vanishes everywhere also. Therefore, for any one-parameter family of geodesics that reach $\bdy$, the norm of the deviation vector $X^a$ is a positive constant along each geodesic.

We now return to $\overM$, and switch to the metric $\tilde g_{ab}$. On $\bdy$, $X^a$ has vanishing norm; being also orthogonal to the geodesic's tangent vector $T^a$, it is proportional to $T^a$ (since orthogonal null vectors are proportional). Without loss of generality, we choose the affine parameter $\lambda$ on each geodesic so that it intersects $\bdy$ at $\lambda=0$; hence, at 
$\lambda=0$, $X^a$ is tangent to $\bdy$. However, by lemma \ref{nullgeo}, $T^a$ is not tangent to $\bdy$. So $X^a=0$. Since this holds for every one-parameter family of geodesics, every connected set of geodesics that reach $\bdy$ intersects it at a point. $\Box$\bigskip

As a warm-up for our main theorem of this subsection, we will now use lemma \ref{point} to prove a version of the Gao-Wald theorem \cite{Gao:2000ga} and a version of the topological censorship theorem \cite{Galloway:1999br}.

\begin{theorem}\label{pointJ}For any point $p\in\bdy$, $\bulkJ^+(p)\cap\bdy=J^+(p)$.\end{theorem}

\emph{Proof:} Clearly $J^+(p) \subset \bulkJ^+(p) \cap\bdy$. Let $t$ be a global time function on $\overM$. Then if $t(q)<t(p)$ we have $q\notin \bulkJ^+(p)$. Therefore, each connected component of $\bdy$ contains some points not in $\bulkJ^+(p)$. Therefore, if 
$\bulkJ^+(p)\cap\bdy\neq J^+(p)$, then $\partial\bulkJ^+(p)\cap\bdy$ includes a hypersurface ${\cal S}$ in $\bdy$ that is not in $J^+(p)$. We will now show that ${\cal S}$ cannot exist.

$\partial \bulkJ^+(p)$ consists of future-directed null geodesics starting at $p$ on which, except at the endpoints, every deviation vector is spacelike and orthogonal to the tangent vector. By lemma \ref{nullgeo}, each such geodesic either lies entirely in $\bdy$ or lies entirely in $\bulk$ except at its endpoints. In particular, the points in ${\cal S}$ must lie on geodesics that are entirely in $\bulk$ except at their endpoints. We thus consider the congruence of geodesics in $\bulk$ starting at $p$. Reversing its direction, every geodesic in this congruence reaches $\bdy$ (at $p$), so the expansion is nowhere negative. Therefore, in the forward direction, its expansion is nowhere positive. Thus the conditions of lemma \ref{point} apply. Hence ${\cal S}$ consists of isolated points, contradicting the fact that it is a hypersurface in $\bdy$. $\Box$

\begin{corollary}\label{wormholes}If $\bdy_1,\bdy_2$ are distinct connected components of $\bdy$, then $\bulkJ^+(\bdy_1)\cap\bdy_2=\emptyset$.\end{corollary}
Corollary \ref{wormholes} rules out traversable wormholes through the bulk connecting different boundary components, and is thus closely related to topological censorship. (A simple argument establishing this can be found in \cite{Freivogel:2005qh}.)

Our goal for the rest of this subsection is generalize Theorem \ref{pointJ} to codimension-two surfaces that are extremal with respect to $g_{ab}$. First, we need two lemmas:

\begin{lemma}\label{Idot}Let ${\cal E}$ be a compact codimension-two submanifold-with-boundary of $\overM$, with boundary ${\cal N}$. Then every point $p\in\partial \bulkJ^+[{\cal E}]$ is on a future-directed null geodesic lying entirely in $\partial \bulkJ^+[{\cal E}]$ that either (1) starts orthogonally from ${\cal E}$ and has no point conjugate to ${\cal E}$ between ${\cal E}$ and $p$, or (2) starts orthogonally from ${\cal N}$, moving away from ${\cal E}$ (i.e.\ $U_aT^a>0$, where $T^a$ is the tangent vector to the geodesic at its starting point, and $U^a$ is a vector at the same point that is tangent to ${\cal E}$, normal to ${\cal N}$, and outward-directed from ${\cal E}$).\end{lemma}

\emph{Proof:} This is a generalization of theorem 9.3.11 in \cite{Wald:1984ai}. Every 
$p\in\partial \bulkJ[{\cal E}]$ lies on a null geodesic starting from ${\cal E}$. If neither condition (1) nor (2) is met, then it can be deformed to a timelike curve and therefore $p\in \bulkI^+[{\cal E}]$. $\Box$

\begin{lemma}\label{normal}Let ${\cal E}$ be a spacelike submanifold-with-boundary of $\overM$ whose restriction to $\bulk$ is extremal with respect to the metric $g_{ab}$. Then ${\cal E}$ intersects $\bdy$ orthogonally, i.e., every normal vector to ${\cal E}$ is tangent to $\bdy$.\end{lemma}

\emph{Proof:} A short calculation shows that, in $\bulk$, the mean curvature $\tilde K^a$ of 
${\cal E}$ with respect to ${\tilde g}_{ab}$ is related to that with respect to $g_{ab}$, $K^a$, as follows:
\begin{equation}
\tilde K^a = \Omega^{-2}K^a + \dim({\cal E})\; \tilde Q^{ab}\partial_b\ln\Omega\,,
\end{equation}
where $\tilde Q^{ab}:={Q^a}_c\; {\tilde g}^{bc}$ and ${Q^a}_c$ is the projector normal to 
${\cal E}$. Since ${\cal E}$ is extremal, $K^a=0$. So
\begin{equation}
\tilde K^2 = \dim({\cal E})^2\; {\tilde Q}^{ab}\; \partial_a\ln\Omega\,\partial_b\ln\Omega\,.
\end{equation}
Since ${\cal E}$ is smooth, $\tilde K^2$ remains finite on $\bdy$, where 
$\ln\Omega\to-\infty$. This requires that every normal vector to 
${\cal E}$ be tangent to $\bdy$. $\Box$

\begin{theorem}\label{surfaceJ}Let ${\cal E}$ be a compact smooth spacelike codimension-two submanifold-with-boundary in $\overM$, whose only boundary is where it intersects $\bdy$, and whose restriction to $\bulk$ is extremal with respect to the metric $g_{ab}$. Then $\bulkJ^+[{\cal E}] \cap\bdy=J^+[{\cal E}\cap\bdy]$.\end{theorem}

\emph{Proof:} The proof is largely a repetition of that of Theorem \ref{pointJ}. Clearly 
$J^+[{\cal E}\cap\bdy]\subset \bulkJ^+[{\cal E}]\cap\bdy$. Let $t$ be a global time function on $\overM$. Since ${\cal E}$ is compact, it has a minimum time $t_{\rm min}$. Clearly if for some point $q\in\bdy$, $t(q)<t_{\rm min}$, then $q\notin \bulkJ^+[{\cal E}]$. Therefore, each connected component of $\bdy$ contains some points not in $\bulkJ^+[{\cal E}]$. Therefore, if $\bulkJ^+[{\cal E}]\cap\bdy\neq J^+[{\cal E}\cap\bdy]$, then $\partial 
\bulkJ^+(m)\cap\bdy$ includes a hypersurface $\Sigma$ in $\bdy$ that is not in $\bulkJ^+[{\cal E}\cap\bdy]$. We will now show that ${\cal S}$ cannot exist.

By lemma \ref{normal}, ${\cal E}$ intersects $\bdy$ orthogonally. Therefore, in lemma \ref{Idot}, the second type of null geodesic in $\partial \bulkJ^+[{\cal E}]$ does not exist. The first type of geodesic forms a codimension-two congruence starting orthogonally from ${\cal E}$ on which, except possibly at the endpoints, every deviation vector is spacelike and orthogonal to the tangent vector. By lemma \ref{nullgeo}, each such geodesic either lies entirely in $\bdy$ or lies entirely in $\bulk$ except at its endpoints. In particular, the points in ${\cal S}$ must lie on geodesics that are entirely in $\bulk$ except where they end. We thus consider the congruence of geodesics in $\bulk$ starting orthogonally from ${\cal E}\cap \bulk$. Since ${\cal E}\cap \bulk$ is extremal, its expansion (with respect to $g_{ab}$) is initially zero. By the null energy condition, its expansion is nowhere positive. Thus the conditions of lemma \ref{point} apply. Hence ${\cal S}$ consists of isolated points, contradicting the fact that it is a hypersurface in $\bdy$. $\Box$\bigskip
 
Note that theorem \ref{pointJ} is a special case of theorem \ref{surfaceJ}, in which we take ${\cal E}$ to be a small (in the metric $\tilde g_{ab}$) hemisphere centered on $p$ and take the limit in which its radius goes to $0$.

\subsection{Spatial regions and causal decompositions}
\label{sec:pspatial}

Let $\Sigma$ be a Cauchy slice of $\bdy$. Given a codimension-zero submanifold of $\Sigma$, let $\regA$ be its interior, $\partial\regA$ its boundary, and $\regAc$ its complement; these three sets do not overlap and cover $\Sigma$. They naturally induce a causal decomposition of the spacetime $\bdy$ into four nonoverlapping regions (except that $J^\pm[\partial\regA]$ both include $\partial\regA$):

\begin{theorem}
\begin{eqnarray}
\domd{\regA} \cup \domd{\regAc} \cup J^+[\partial\regA]\cup J^-[\partial\regA] &=& \bdy
\label{covers}\\
D[\regA]\cap D[\regAc]=\domd{\regA} \cap J^\pm[\partial\regA] = \domd{\regAc} \cap J^\pm[\partial\regA] 
&=& \emptyset \label{nonoverlapping1}\\
J^+[\partial\regA]\cap J^-[\partial\regA]&=&\partial\regA\,.\label{nonoverlapping2}
\end{eqnarray}
\label{decomposition}\end{theorem}

\emph{Proof:} \eqref{nonoverlapping1} and \eqref{nonoverlapping2} are obvious from the definitions. 

We now prove \eqref{covers}. Suppose a point $p\in J^+[\Sigma]$ is not in any of the four regions. Each inextendible causal curve through $p$ intersects $\Sigma$ exactly once, but not in $\partial\regA$ (else $p\in J^+[\partial\regA]$). Nor can all such curves intersect it in $\regA$ (else $p\in D[\regA]$) or $\regAc$ (else $p\in D[\regAc]$). So some must intersect $\Sigma$ in $\regA$ and others in $\regAc$. Let $\lambda_1$ be in the first set and $\lambda_2$ in the second. Join $\lambda_1$ and $\lambda_2$ at $p$ to make a continuous curve $\lambda$ from $\regA$ to $\regAc$. Now, in any globally hyperbolic spacetime there exists a global timelike vector field; its integral curves can be used to construct a continuous map $f$ from $J^+(\Sigma)$ to $\Sigma$. $f(\lambda)$ is a continuous curve in $\Sigma$ from $\regA$ to $\regAc$. There therefore exists a point $q\in\lambda$ such that $f(q)\in\partial\regA$, and therefore $q\in I^+[\partial\regA]$. Since $p\in J^+(q)$, $p\in J^+[\partial\regA]$, which is a contradiction. 
$\Box$\bigskip

Now let $\extr$ be a surface in $\overM$ that satisfies the conditions of theorem \ref{surfaceJ} and is spacelike-homologous to $\regA$. The precise meaning of the latter condition is as follows: There exists a Cauchy slice $\tilde\Sigma$ for $\overM$ such that $\tilde\Sigma\cap\bdy=\Sigma$, containing a codimension-zero submanifold with boundary $\regA\cup\extr$; we call its interior $\homsurfA$. Since $\tilde\Sigma$ is itself a manifold-with-boundary (namely $\tilde\Sigma\cap\bdy$), one has to be careful about the definitions of ``interior'' and ``boundary'' for a submanifold. We mean ``interior'' in the sense of point-set topology; thus $\homsurfA$ includes $\regA$ but not $\extr$. The ``boundary'' can be either in the sense of ``submanifold-with-boundary'' (which is what we call $\partial\homsurfA$), or in the sense of point-set topology. In the latter sense, the boundary is just $\extr$.\footnote{ The point-set-topology boundary can be shown to equal the ``edge'' of the submanifold, in the sense used in the general-relativity literature (see e.g.\ \cite{Wald:1984ai}).} As with $\regA$, we define $\homsurfA^c:=\tilde\Sigma\setminus(\homsurfA\cap\extr)$. To summarize, in parallel to the decomposition of $\Sigma$ into $\regA$, $\regA^c$, and $\partial\regA$, we have a decomposition of $\tilde\Sigma$ into $\homsurfA$, $\homsurfA^c$, and $\extr$. Furthermore, $\homsurfA\cap\bdy=\regA$, $\homsurfA^c\cap\bdy=\regA^c$, and $\extr\cap\bdy=\partial\regA$.

We can now apply theorem \ref{decomposition} to obtain a decomposition of $\overM$ into the four spacetime regions $D[\homsurfA]$, $D[\homsurfA^c]$, $J^\pm[\extr]$. The central result of this section is that this decomposition reduces on the boundary precisely to its decomposition into $D[\regA]$, $D[\regA^c]$, and $J^\pm[\partial\regA]$:

\begin{theorem}\label{compatible}
\begin{subequations}
\begin{align}
\bulkD[\homsurfA] \cap\bdy&=\domdA
\label{D(R)}\\
\bulkD[\homsurfA^c]\cap\bdy&=\domdAc
\label{D(Rc)}\\
\bulkJ^\pm[\extr]\cap\bdy&=J^\pm[\entsurf]
\label{Ipm}
\end{align}
\end{subequations}
\end{theorem}
\emph{Proof:} Equation \eqref{Ipm} is Theorem \ref{surfaceJ} (and its time reverse). Using Theorem \ref{decomposition} both in $\bdy$ and in $\overM$ to take the complement of both sides, we have
\begin{equation}
\left(\bulkD[\homsurfA]\cap\bdy\right)\cup\left(\bulkD[\homsurfAc]\cap\bdy\right) = \domdA\cup \domdAc \,.
\end{equation}
Lemma \ref{domaincontained} then implies \eqref{D(R)}, \eqref{D(Rc)}. $\Box$\bigskip

Theorem \ref{compatible} immediately implies that $\extr$ is outside of causal contact with $D[\regA]$ and $D[\regAc]$, as required by field-theory causality.

The spacelike-homology condition raises the following practical question: Given a codimension-one submanifold of $\overM$ with boundary $\regA\cup\extr$, under what circumstances is it contained in a Cauchy slice? Obviously, it must be acausal. However, this is not sufficient; for example, a spacelike hypersurface in Minkowski space that approaces null infinity is not contained in a Cauchy slice. The following lemma, which will also be needed in theorem \ref{wedgeobservable}, shows that compactness is a sufficient additional condition. (This lemma applies in any globally hyperbolic spacetime.)

\begin{lemma}If $R$ is a compact acausal set, then there exists a Cauchy slice containing it.\label{Cauchyexistence}\end{lemma}

\emph{Proof:} Let $t \in {\mathbb R}$ be a global time function, and define $t_{\rm max}:=\max_R(t)$, $t_{\rm min}:=\min_R(t)$ (these exist since $R$ is compact). Define $\Upsilon:=\{p:t>t_{\rm max}\}$ and $\Upsilon':= \Upsilon \cup \bdyI^+[R]$. Define
\begin{equation}
\Sigma:= \partial \Upsilon'= \left(\partial \Upsilon\setminus \bdyI^+[R]\right)\cup \left( \partial\bdyI^+[R]\setminus \Upsilon\right).
\end{equation}
$\partial \bdyI^+[R]$ contains $R$, and $\Upsilon\cap R=\emptyset$, so $R\subset\partial \bdyI^+[R]\setminus \Upsilon\subset\Sigma$. Next we show that $\Sigma$ is achronal. The maximum value of $t$ on $\Sigma$ is $t_{\rm max}$, so there can be no future-directed timelike curve from $\partial \Upsilon$ to $\Sigma$. Further $\partial \bdyI^+[R]$ is itself achronal. Finally, if there is a future-directed timelike curve from $p\in\partial \bdyI^+[R]$ to $q\in\partial\Upsilon$, then $q\in \bdyI^+[R]$ and hence $q\not\in\Sigma$. So $\Sigma$ is achronal. 

Next, we show that every inextendible future-directed timelike curve intersects $\Sigma$. On such a curve, $t$ increases monotonically and continuously from $-\infty$ to $+\infty$. For $t\le t_{\rm min}$, the curve is not in $\Upsilon'$; for $t>t_{\rm max}$, it is. Therefore for some value of $t$ it intersects $\Sigma$. 

While $\Sigma$ is achronal, it is not quite a Cauchy slice (in the sense used in this paper) because it is not acausal. However, since $R$ is acausal, $\Sigma$ can be deformed outside of $R$ to be acausal. $\Box$

\begin{theorem}\label{wedgeobservable}Let $\Sigma'$ be a Cauchy slice for $\bdy$ and $\regA'\subset\Sigma'$ a region such that $\regA'\cup\partial\regA'$ is compact and $D[\regA']=D[\regA]$. Then $\regA'$ is spacelike-homologous to $\extr$.\end{theorem}

\emph{Proof:} Since $\extr$ and $\regA'\cup\partial\regA'$ are both compact, $\extr\cup\regA'$ is compact as well. (Recall that $\partial\regA'=\partial\regA\subset\extr$.) $\extr$ and $\regA'$ are acausal, since each sits on a Cauchy slice. Furthermore, by theorems \ref{decomposition} and \ref{surfaceJ}, there are no causal curves connecting them; hence $\extr\cup \regA'$ is acausal. Therefore, by theorem \ref{Cauchyexistence}, there is a Cauchy slice $\tilde\Sigma'$ containing both $\extr$ and $\regA'$.

Choosing a global timelike vector field on $\overM$, its integral curves define a diffeomorphism $f:\tilde\Sigma\to\tilde\Sigma'$. Let $\homsurfA':=f(\homsurfA)$. Since $\extr$ is contained in both $\Sigma$ and $\Sigma'$, $f(\extr)=\extr$. Since every timelike curve in $\domdA$ intersects $\Sigma$ in $\regA$ and $\Sigma'$ in $\regA'$, $f(\regA)=\regA'$. So $\homsurfA':=f(\homsurfA)$ is a region in $\tilde\Sigma'$ with $\partial\homsurfA'=\regA'\cup\extr$. (Strictly speaking, we also need to define a new Cauchy slice for $\bdy$, $\Sigma'':=\tilde\Sigma'\cap\bdy$, and to consider $\regA'$ to be a region in $\Sigma''$, since the equality  $\Sigma''=\tilde\Sigma'\cap\bdy$ is part of the definition of the spacelike homology condition.) $\Box$\bigskip

Theorem \ref{wedgeobservable} shows that the HRT formula gives the same value for the entanglement entropy of $\regA$ and $\regA'$, as required by field-theory causality.

\section{Discussion}
\label{sec:discuss}

The main result of this paper, Theorem \ref{compatible}, shows that the HRT prescription for computing holographic entanglement entropy \cite{Hubeny:2013gta} is consistent with the requirements of field theory causality. As we have explained with various simple examples and gedanken experiments in \S\ref{sec:gexpt}, the result was in no way \emph{a priori} obvious, since there are several marginal cases where arbitrarily small deformation of the bulk extremal surface would place it in causal future of a boundary deformation which however cannot affect the entanglement entropy. With the primary result at hand, we now take stock of the various physical consequences it implies for holographic field theories. 

\paragraph{Causality constraints on holography:}
Let us start by asking what we can learn about holography from causality considerations. Recall that we proved our result for extremal surfaces in the context of two-derivative theories of gravity satisfying the null energy condition. This was crucial for us to be able to use the Raychaudhuri equation in order to ascertain properties of null geodesic congruences. Thus the domain of validity of our statements was strong coupling in a planar (large-$N$) field theory. This translates to demanding a macroscopic spacetime with $\ell_s \ll \ell_{\rm AdS}$ in a perturbative string  ($g_s \ll1$) regime. Lets see what happens as we move away from this corner of moduli space.

Firstly, consider classical stringy corrections which we can encapsulate in an effective higher-derivative theory of gravity. In such a theory, as long as higher-derivative operators are suppressed by powers of $\ell_s$, our conclusions will hold, since the dominant effect will come from the leading two-derivative Einstein-Hilbert term in the bulk.  When the  higher-derivative operators are unsuppressed we have little to say  for two reasons: 
(a) the holographic entanglement prescription so far is only  given for  static situations (or with time reversal symmetry) \cite{Dong:2013qoa,Camps:2013zua} and (b) even assuming the covariant generalizations, one is stymied by the absence of clean statements regarding dynamics of null geodesic congruences (even for example in Lovelock theories).\footnote{  The family of $f(R)$ theories can be brought to heel, since here we can map the theory to Einstein-Hilbert via a suitable Weyl transformation. Causality constraints can be discerned here so long as the Weyl transformation (which is non-linear in the curvature) is well-behaved.} One could, however, use the causality constraint to rule out certain higher-derivative theories from having unitary relativistic QFT duals (see e.g.\ \cite{Erdmenger:2014tba}); this is similar in spirit to the recent discussions on causality constraints on the three-graviton vertex \cite{Camanho:2014apa}.

Turning next to $1/N$, or bulk quantum corrections, while we have less control in general, we can make some observations about the leading $1/N$ correction which has been proposed to be given by the entanglement of bulk perturbative quantum fields across $\extr$ \cite{Faulkner:2013ana}. Since the bulk theory itself is causal, it follows that entanglement across the extremal surface satisfies the desired causality conditions.

\paragraph{Does causality prove the HRT conjecture?:}  One intriguing possibility given, the importance of the causality, is whether we can use it to constrain the location of the extremal surface in the bulk, and thus prove the HRT conjecture.\footnote{ We thank Vladimir Rosenhaus for inspiring us to think through this possibility.} Unfortunately, causality alone is not strong enough to pin down the location of the extremal surface. What we can say is that the extremal surface $\extr$ has to lie inside the causal shadow $\shadow_{\entsurf}$. In a generic asymptotic AdS spacetime, for a generic region $\regA$, the casual shadow is a codimension-zero volume of the bulk spacetime $\bulk$. It is only in some very special cases that we  zero in on a single bulk codimension-two surface uniquely (e.g., spherical regions in pure AdS or in the eternal Schwarzschild-AdS black hole).\footnote{ The examples are all cases where, by a suitable choice of conformal frame, the extremal surface can be mapped onto the bifurcation surface of a static black hole. The black funnel and droplet solutions (see \cite{Marolf:2013ioa} for a review) provide nontrivial examples, cf., \cite{Hubeny:2013gba}.}

\paragraph{Causality constraints on other CFT observables:}  
Our discussion has exclusively focused on the causality properties of a particular non-local quantity in the field theory, namely the entanglement entropy. However, causality places restrictions on other physical observables we can consider on the boundary as well. For instance, correlation functions of (time-ordered) local operators, Wilson loop expectation values, etc., should all obey appropriate constraints which we can infer from basic principles. Indeed, this can be shown to be the case, for example, for correlation functions, by considering the fact that the bulk computation involves solving a suitable boundary initial value problem for fields in the bulk, which can be checked to manifestly satisfy causality. 

However, this is less clear when we approximate, say, two point functions of heavy local operators using the geodesic approximation \cite{Balasubramanian:1999zv}. Similar issues arise for the semi-classical computation of Wilson loop expectation values \cite{Maldacena:1998im,Rey:1998ik} using the string worldsheet area. In these cases, one generically encounters some tension between the use of extremal surfaces---geodesics, two-dimensional worldsheets, etc.---for the bulk computation, and field theory expectations regarding causality (cf., \cite{Louko:2000tp} for an earlier discussion of this issue). Indeed, it appears that codimension-two extremal surfaces are special in this regard, for we can rely  on the boundary of the entanglement wedge being generated by a codimension-one null congruence, and thus apply the Raychaudhuri equation. Understanding the proper application of the WKB approximation for other observables is an interesting question; we hope to report upon in the near future \cite{Headrick:2014gf}.

\paragraph{Entanglement wedges:}
One of the key constructs in our presentation, naturally associated with a given boundary region $\regA$, has been the entanglement wedge $\EWA$.  This is the domain of dependence of the homology surface $\homsurfA$ (recall that $\homsurfA$ forms a part of a Cauchy surface which interpolates between $\regA$ and $\extr$).  Equivalently, it comprises the set of spacelike-separated points from $\extr$ which is connected to $\regA$, one of the four regions in the natural decomposition of the bulk spacetime.

Given $\regA$, one might ask how unique this decomposition is.  Since $\EWA$ is a causally-defined set, its specification only requires the specification of the (oriented) extremal surface $\extr$ (possibly consisting of multiple components when so required by the homology constraint).  The prescription for constructing the null boundary of $\EWA$ is unambiguous: simply to follow all null normals (emanating from $\extr$ in the requisite direction, towards $\domdA$) until they encounter another generator (i.e.\ a crossover seam) or a caustic.  
However, there is a possibility that the extremal surface itself is not uniquely determined from $\regA$.  This happens when multiple (sets of) extremal surfaces satisfy \eqref{SAdef} but have the same area.  Since entanglement entropy itself cares only about the area, the HRT (as well as RT and maximin) prescription is to take {\it any} of these.  However, which we take {\it does} matter for the entanglement wedge.  We propose that, just as for the extremal surfaces, in such cases we may have {\it multiple} entanglement wedges $\EWA$ associated to the same boundary region $\regA$.  

The most ``obvious'' class of examples where this can happen is the case of $\regA$ consisting of multiple regions or in higher dimensions where the entangling surface $\entsurf$ consists of multiple disjoint components.  As we vary the parameters describing the configuration, the extremal surfaces involved typically exchange dominance, so at some point their areas must agree.  Applying continuity from both sides, at the transition point, both entanglement wedges should be naturally associated with $\regA$.
However, in complicated states, there can actually be multiple extremal surfaces even for when $\regA$ and $\entsurf$ are both connected.  In such cases, we could have candidate entanglement wedges which are proper subsets of (rather than merely overlapping with) other candidate entanglement wedges.  

\begin{figure}
\begin{center}
\includegraphics[width=4in]{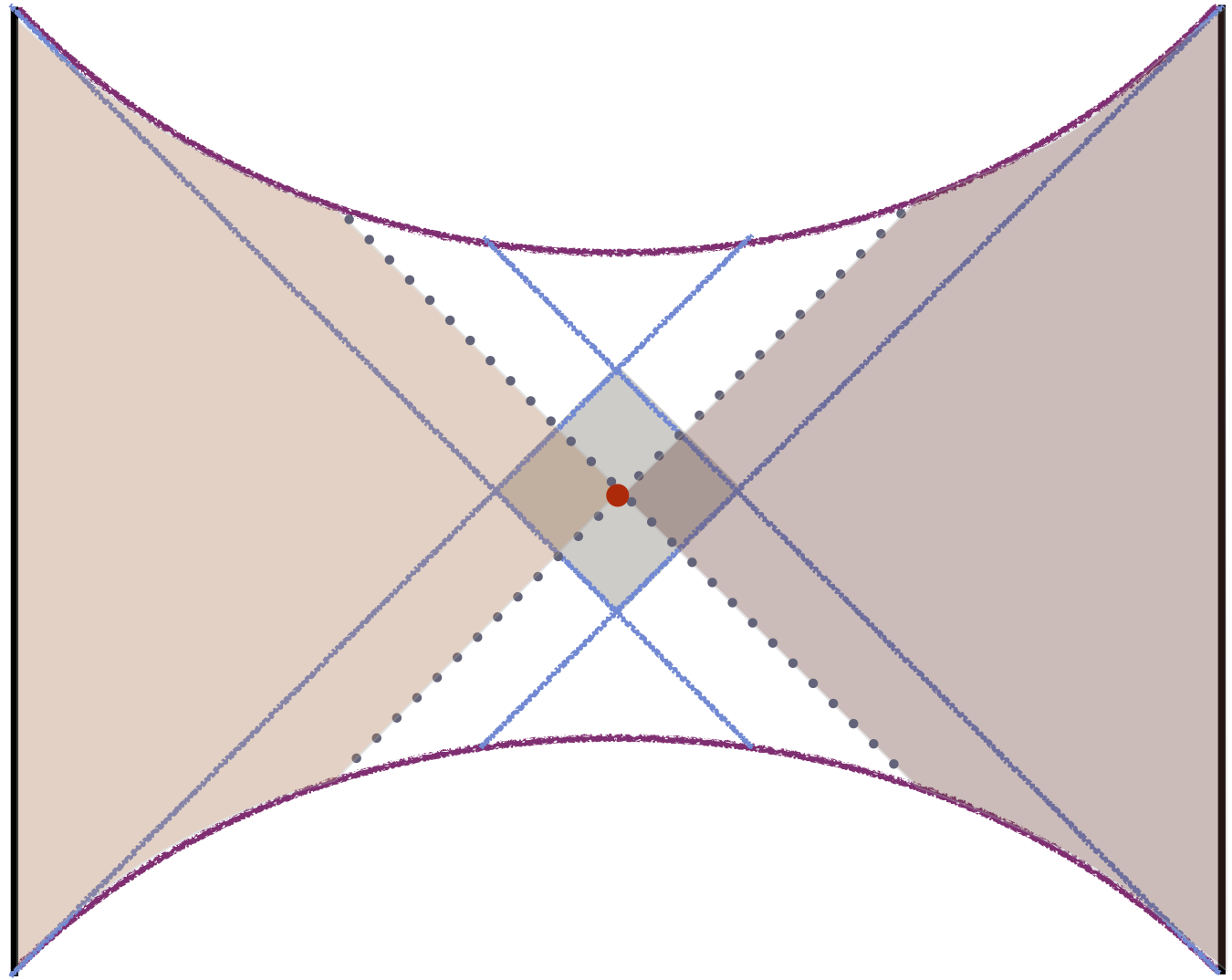}
\setlength{\unitlength}{0.1\columnwidth}
\begin{picture}(0.3,0.4)(0,0)
\put(0.2,3){\makebox(0,0){CFT$_R$}}
\put(-7.15,3){\makebox(0,0){CFT$_L$}}
\put(0.,2){\makebox(0,0){$\regA$}}
\put(-3.45,2.9){\makebox(0,0){$\shadow$}}
\put(-5.5,2.5){\makebox(0,0){$\EWAc$}}
\put(-1.5,2.5){\makebox(0,0){$\EWA$}}
\end{picture}
\caption{
Sketch of Penrose diagram for a symmetric Vaidya-\SAdS{}  geometry obtained by imploding null shells to the past and future from both boundaries now displaying the entanglement wedges and the causal shadow region, with $\regA$ being a full Cauchy surface for CFT$_R$.
}
\label{f:VSAdSewedge}
\end{center}
\end{figure}
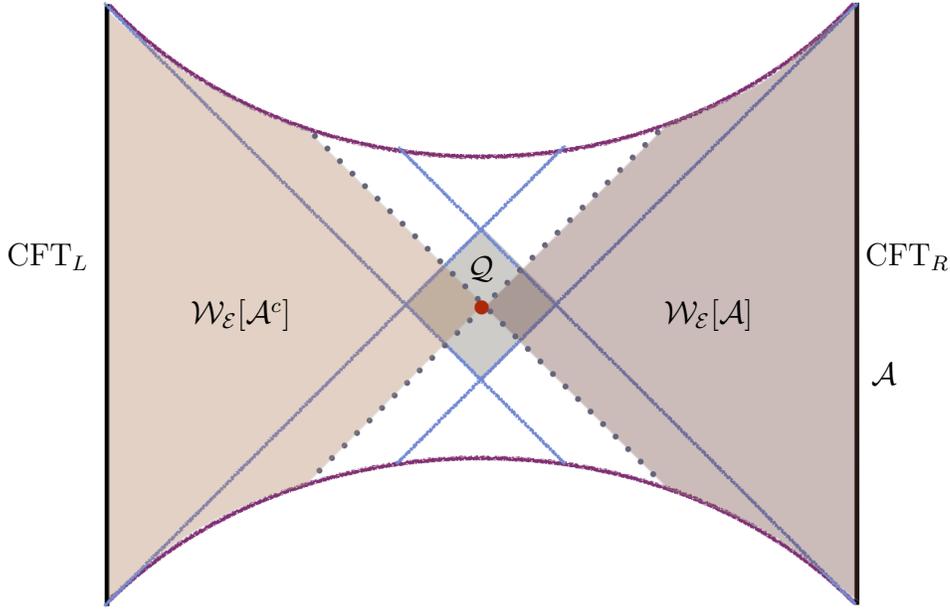

It is also interesting to note that the decomposition of the bulk into four spacetime regions causally defined from $\extr$ need not coincide with the bulk decomposition defined from ${\cal E}_{\regAc}$, despite there being a unique boundary decomposition defined from $\entsurf$.  For pure states, where the homology constraint trivializes and we have  $\extr = {\cal E}_{\regAc}$, we can write the bulk decomposition equivalently with respect to both $\regA$ and $\regAc$,
\begin{equation}
\bulk = \EWA \cup \EWAc \cup \bulkJ^+[\extr] \cup \bulkJ^-[\extr]
\label{bulkdecomp}
\end{equation}	
which is directly analogous to the boundary decomposition \eqref{bdy4d}.  However, for mixed states, where typically $\extr \ne {\cal E}_{\regAc}$, the decomposition \eqref{bulkdecomp} is {\it not} true;\footnote{ Note however that if we purify a mixed state by additional boundaries, such as in the deformed eternal black hole example illustrated in Fig.~\ref{f:VSAdSewedge}, then the decomposition \eqref{bulkdecomp} does hold.} instead the correct decomposition should replace $\EWAc$ with the bulk domain of dependence of the complement of $\homsurfA$ within the bulk Cauchy slice ${\tilde \Sigma}$, or more precisely 
$\bulkD[{\tilde \Sigma} \backslash \homsurfA \backslash \extr]$.

\paragraph{Dual of $\rhoA$?}
Within the class of CFTs and states with a geometrical holographic dual, it has often been asked,\footnote{ In recent years this question has been invigorated by e.g.\ \cite{Bousso:2012sj,Czech:2012bh}.  
} for a given region $\regA$, what is the bulk ``dual" of the reduced density matrix $\rhoA$.  One way to formulate what one means by this is as follows: suppose we fix $\rhoA$ and vary over all compatible density matrices for the full state $\rho$.  What is the maximal bulk spacetime region which coincides for all such $\rho$'s?  By ``coinciding bulk regions'' one means having the same geometry, i.e.\ the same bulk metric modulo diffeomorphisms. Another way to define the dual of $\rhoA$ is to ask what is the maximal bulk region wherein we can uniquely determine the bulk metric (again modulo diffeomorphisms).
In fact there  are several (generally distinct) bulk regions that might be naturally associated with the density matrix;  in nested order:
\begin{itemize}
\item The bulk region that $\rhoA$ is {\it sensitive to}; in other words, regions wherein a deformation of the metric affects $\rhoA$.\footnote{ 
In fact there is a further subdivision here based on whether {\it any }geometrical deformation of the metric should change $\rhoA$ or merely whether there should exist {\it some} deformation of the metric which changes $\rhoA$.
We thank Mark Van Raamsdonk for discussions on this issue.
}
\item The bulk region that $\rhoA$ {\it determines}, i.e.\ where we can uniquely reconstruct all the components of the metric (up to diffeomorphisms).
\item The bulk region that $\rhoA$ {\it affects}, i.e.\ where by changing $\rhoA$ one can change the bulk metric.  
\end{itemize}

Here we focus on the second case, following \cite{Bousso:2012sj,Czech:2012bh}. Based on lightsheet arguments, the authors of \cite{Bousso:2012sj} proposed the causal wedge as the correct dual.  On the other hand, \cite{Czech:2012bh}, as well as \cite{Hubeny:2012wa, Wall:2012uf}, argued that the requisite region should contain more than the causal wedge.  In particular, \cite{Czech:2012bh} presented a number of criteria that such a region should satisfy, and explored several possibilities, most notably the region they denoted ${\hat w}(D_A)$ which corresponds to the bulk domain of dependence of the spacetime region spanned by all codimension-two extremal surfaces anchored within $\domdA$.  If every point of $\homsurfA$ lies on at least one of these, then this region coincides with our entanglement wedge $\EWA$.  On the other hand, as \cite{Czech:2012bh} pointed out, there may be ``holes'' in such a set, i.e., regions of $\homsurfA$ which do not lie along any least-area extremal surface anchored on a given region $\regA' \subset \regA$.\footnote{ 
The example given in \cite{Czech:2012bh} involves a region through which traversing surfaces are not the smallest-area ones anchored on the given region, 
but a simpler physical example would be a point sufficiently close to an event horizon of an eternal spherical black hole, with $\regA = \Sigma$ of one side as considered in \S\ref{sec:gexpt}.}

We propose that, since the most ``natural'' causal set associated with $\rhoA$ from the bulk point of view is the entanglement wedge, this is indeed the most appropriate region to be identified with the ``dual'' of the reduced density matrix $\rhoA$ (even in the presence of such entanglement ``holes''). In this context, we should note that we can strip away the rest of the boundary spacetime, and consider the field theory just on $D[\regA]$, which is a globally hyperbolic spacetime in its own right, in the state $\rho_\regA$. Whether this state in general admits a holographic description is not known, but, if it does, then a natural candidate would seem to be the entanglement wedge: this is, in its own right, a globally hyperbolic, asymptotically AdS spacetime, whose conformal boundary (according to theorem \ref{compatible}) is precisely $D[\regA]$, and the area of whose edge $\extr$ gives the entropy of $\rho_\regA$.

Here the word ``natural'' should be qualified, especially in light of the arguments in \cite{Hubeny:2012wa} that the causal wedge $\CWA$ is a natural bulk codimension-zero region associated with $\regA$.  The latter can be obtained more minimally: it suffices to know the causal structure of the bulk to define $\CWA$.  On the other hand, the density matrix clearly encodes much more than the bulk causal structure, since at least it knows the entanglement entropy (as well as entanglement entropies of all subregions, apart from other observables).  Since, in the bulk, the corresponding extremal surface is defined only once we know the bulk geometry, the entanglement wedge $\EWA$ it defines is a less minimal construct that the causal wedge $\CWA$.  Nevertheless, once $\extr$ is identified, the rest of the bulk construction of the entanglement wedge is purely causal, and therefore defined fully robustly for any time-dependent asymptotically AdS spacetime.

\begin{figure}
\begin{center}
\includegraphics[width=4in]{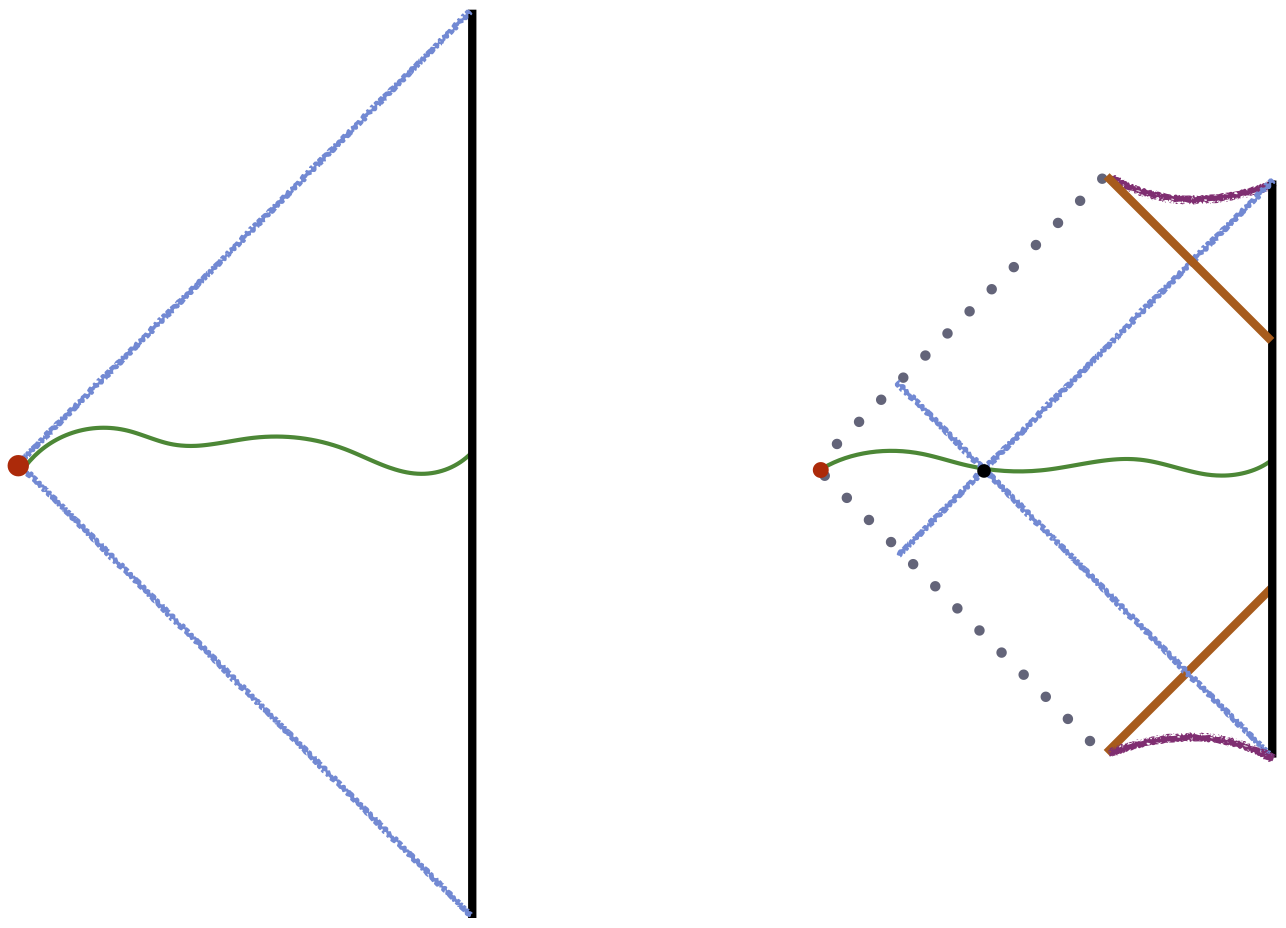}
\setlength{\unitlength}{0.1\columnwidth}
\begin{picture}(0.3,0.4)(0,0)
\put(0,2.35){\makebox(0,0){$\regA$}}
\put(-4.1,2.35){\makebox(0,0){$\regA$}}
\put(-5.5,2.65){\makebox(0,0){$\color{olivegreen}{\homsurfA}$}}
\put(-1,2.55){\makebox(0,0){$\color{olivegreen}{\homsurfA}$}}
\put(-6.9,2.35){\makebox(0,0){$\color{red}{\extr}$}}
\put(-2.8,2.35){\makebox(0,0){$\color{red}{\extr}$}}
\end{picture}
\caption{
Left: Exterior AdS-Schwarzschild solution, dual to a deconfined thermal state on $S^{d-1}$. The extremal surface for the entire boundary (red dot) coincides with the bifurcation surface and the causal information surface. Right: Vaidya solution with an ingoing null shell that reaches the boundary at $t<0$ and an outgoing one that leaves it at $t>0$ (brown); the geometry between the shells is unchanged, but the past and future event horizons (blue) have moved closer to the boundary, leaving the extremal surface (red dot) hidden behind them. The entanglement wedge in both cases is the entire spacetime (with a homology surface shown in green), while the causal wedge in the right figure is just the part outside of the event horizons. (The causal information surface is shown as the black dot.)
}
\label{f:wedgeargument}
\end{center}
\end{figure}
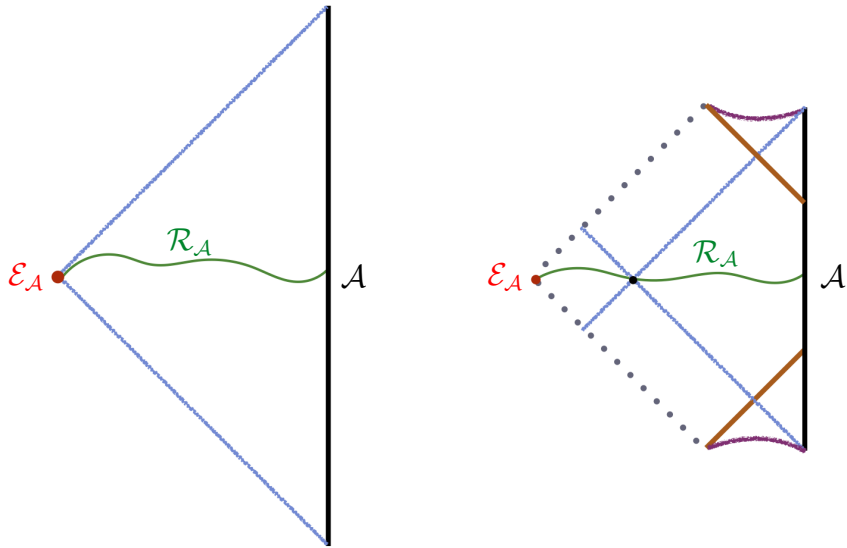

The statement that the entanglement wedge is the natural dual of the reduced density matrix (which implies that the boundary observer in $\domdA$ can learn about the bulk geometry in the entire $\EWA$) has a profound consequence. We have shown that the extremal surface $\extr$ has to lie in the causal shadow.  This set can however be quite large, and so $\extr$ can lie very deep inside the bulk (as indicated by the shaded region in Fig.~\ref{f:VSAdSewedge}). In fact, a simple example supports the idea that the entanglement wedge represents the state in such a case (see Fig.~\ref{f:wedgeargument}). We start with a deconfined thermal state at $t=0$ on a single ${\bf S}^{d-1}$, represented holographically by the exterior Schwarzschild-AdS solution. We add an outgoing null shell that reaches the boundary at $t<0$ and an ingoing one that leaves it at $t>0$. At $t=0$ we still have the thermal state. The bulk solution is also unchanged between the past and future shells. However, these shells move the singularity and therefore have the effect of bringing the future and past event horizons closer to the boundary, leaving the previous bifurcation surface hidden behind both horizons. While this surface is no longer the bifurcation surface of a global Killing vector, it remains the extremal surface whose area gives the entropy of the state of the field theory on the right boundary. Presumably the holographic description of the state extends all the way down to this extremal surface, as it does in the absence of the shells, and thus consists of the entire entanglement wedge.

Another (related) example where the separation between entanglement wedge and causal wedge is particularly striking is the eternal (two-sided) black hole deformed by many shocks considered in  \cite{Shenker:2013pqa,Shenker:2013yza}.  The Einstein-Rosen bridge is highly elongated and the extremal surface probably lies somewhere in the middle of it---so that the entanglement wedge for the entire right boundary is substantially larger than the causal wedge, which in this case is simply the right exterior (domain of outer communication) of the black hole.  So not only does the entanglement wedge penetrate arbitrarily close to the curvature singularity, it also contains a substantial part of the spacetime far beyond the black hole horizon!

\acknowledgments 

We would like to thank Hong Liu, Juan Maldacena, Don Marolf, Steve Shenker, Mark Van Raamsdonk,  and Aron Wall for discussions. MH, VH, and MR would like to thank University of Amsterdam for hospitality during the initial stages of this project. VH and MR would also like to acknowledge the hospitality of Brandeis University, MIT, Cambridge University, and the University of British Columbia, Vancouver during various stages of this project.  We would like to thank the Aspen Center for Physics (supported by the National Science Foundation under Grant 1066293) for their hospitality during both intermediate and concluding stages of this project.

MH is supported in part by the National Science Foundation under CAREER Grant No.\ PHY10-53842. VH and MR are supported in part by the Ambrose Monell foundation, by the FQXi  grant ``Measures of Holographic Information" (FQXi-RFP3-1334) and  by the STFC Consolidated Grant ST/J000426/1. AL is supported in part by DOE grant DE-SC0009987.
MR is supported by the European Research Council under the European Union's Seventh Framework Programme (FP7/2007-2013), ERC Consolidator Grant Agreement ERC-2013-CoG-615443: SPiN (Symmetry Principles in Nature).

\providecommand{\href}[2]{#2}\begingroup\raggedright\endgroup

\end{document}